\documentclass[aps,pre,twocolumn,showpacs]{revtex4}

\usepackage{amsmath}
\usepackage{amssymb}
\usepackage{graphicx}
\usepackage{dcolumn}
\usepackage{bm}

\newcommand{\al}{\alpha}
\newcommand{\om}{\omega}

\newcommand{\tk}{\widetilde{k}}
\newcommand{\tom}{\widetilde{\omega}}
\newcommand{\tal}{\widetilde{\alpha}}

\newcommand{\prt}{\partial}

\newcommand{\ou}{{\overline{u}}}
\newcommand{\ox}{{\overline{x}}}
\newcommand{\sn}{\mathrm{sn}}

\begin{document}

\title{
Dispersive shock wave theory for nonintegrable equations}

\author{ A. M. Kamchatnov}
\affiliation{Institute of Spectroscopy,
Russian Academy of Sciences, Troitsk, Moscow, 108840, Russia}
\affiliation{Moscow Institute of Physics and Technology, Institutsky
lane 9, Dolgoprudny, Moscow region, 141700, Russia}

\date{\today}

\begin{abstract}
We suggest a method for calculation of parameters of dispersive shock waves
in framework of Whitham modulation theory applied to non-integrable wave
equations with a wide class of initial conditions corresponding to
propagation of a pulse into a medium at rest. The method is based on
universal applicability of Whitham's `number of waves conservation law' as
well as on the conjecture of applicability of its soliton counterpart to
the above mentioned class of initial conditions which is substantiated by
comparison with similar situations in the case of completely integrable wave
equations. This
allows one to calculate the limiting characteristic velocities of the Whitham
modulation equations at the boundary with the smooth part of the pulse whose
evolution obeys the dispersionless approximation equations. We show that
explicit analytic expressions can be obtained for laws of motion of the edges.
The validity of the method is confirmed by its application to similar situations
described by the integrable Korteweg-de Vries (KdV) and nonlinear
Schr\"{o}dinger (NLS) equations and by comparison with the results
of numerical simulations for the generalized KdV and NLS equations.
\end{abstract}


\maketitle

\section{Introduction}
\label{intro}

It is well known that if one neglects effects of dissipation and dispersion, then the
theory of propagation of nonlinear waves suffers from breaking singularity developed
at some finite moment of time $t_b$ after which a formal solution of nonlinear wave
equations becomes multi-valued. The classical approach to resolution of this problem
is based physically on taking into account dissipation effects resulting in formation
of shock waves instead of the non-physical multi-valued solutions. Since in practice the
dissipation is often very small and, consequently, the width of shocks is very small, too,
such shocks can be treated as surfaces of discontinuities of physical parameters whose
values at both sides of such a surface are related by {\em jump conditions} derived
from conservation of mass of the fluid, its momentum and energy \cite{LL-6,CF-50}.
As a result, one arrives at remarkably powerful theory which has found numerous
applications in science and technology.

However, in many physical systems the dissipative effects are relatively small
compared with dispersion effects. The well-known example of such a situation is
given by ``undular bores'' in shallow water waves
theory (see, e.g., \cite{bl-54}). In this case, interplay of relatively strong
dispersion and weak dissipation leads to formation of a wide stationary oscillatory
structure instead of a standard shock discontinuity.
The generality of this situation was underlined
by Sagdeev \cite{sagdeev} who considered similar structures in plasma physics and
quantitative asymptotic theory of such undular bores was developed by
Johnson~\cite{johnson-70} in framework of the Korteweg-de Vries-Burgers equation
incorporating both damping and dispersion.

If the damping effects are measured by some parameter (say, the viscosity coefficient)
$\nu$, then the asymptotic stationary stage is developed at very large time
$t\gtrsim\nu^{-1}$, hence for considerable period $t\ll\nu^{-1}$ these damping effects
can be neglected and the wave evolution is essentially non-stationary.
In this case, short-wavelength nonlinear oscillations are
generated after the wave breaking moment instead of a classical viscous shock
discontinuity. Since such a non-stationary oscillatory structure describes transition
between two different states of smooth flow, it is usually called {\em dispersive shock wave}
(DSW). The simplest and, apparently, most powerful theoretical approach to description
of DSWs was formulated by Gurevich and Pitaevskii \cite{gp-73} in framework
of the Whitham theory of modulation of nonlinear waves \cite{whitham-65,whitham-74}
which was based on large difference between scales of the
wavelength of nonlinear oscillations within DSW and the size of the whole DSW.
Further development of nonlinear physics demonstrated the generality of this phenomenon
and now they have been observed in a number of physical
situations (see, e.g., review article \cite{eh-16} and references therein).

Due to the above mentioned
difference of time scales, the modulation parameters change slowly along the DSW and
their evolution is governed by the Whitham equations which can be obtained by
averaging of the conservation laws of the wave equation under consideration over
fast oscillations within the DSW. Although this approach is very general, its
practical applicability depends crucially on mathematical properties of the nonlinear
wave equations which describe the evolution of the wave. The great achievement of
Whitham was that in the case of waves whose evolution is described by the
Korteweg-de Vries (KdV) equation he succeeded in transformation of the modulation
equations to the so-called diagonal Riemann form. Gurevich and Pitaevskii
used just this form of modulation equations in their seminal paper \cite{gp-73}.
It became clear later \cite{ffml-80}, that Whitham's diagonalization of the
modulation equations was possible due to the special property, discovered in
Ref.~\cite{ggkm-67}, of {\em complete integrability} of the KdV equation. Development
of the finite-gap integration method \cite{lax-74,nov-74} as well as of the methods
of derivation \cite{ffml-80,krich-88} and solution \cite{tsarev,dn-93} of the Whitham
equations permitted one to extend the Gurevich-Pitaevskii
approach to a number of other completely integrable equations of physical interest
(see, e.g., \cite{eh-16}). Complemented by the perturbation theory, derived for
the KdV equation with small viscosity term added \cite{fml-84,akn-87,gp-88,mg-95}
and later generalized to a wide class of perturbed completely integrable equations
in Ref.~\cite{kamch-04}, this theory has found many important applications.

On the contrary, development of the general Whitham method of modulations, not restricted
to completely integrable equations, was much slower
and its progress was quite limited. Apparently, the first general statement was made by
Gurevich and Meshcherkin \cite{gm-84}, who proposed the condition which replaces the
``jump conditions'' known in the theory viscous shocks. This condition is formulated
in terms of Riemann invariants of the hydrodynamic system obtained in the dispersionless
limit of the original nonlinear wave equations under consideration. Typically, wave
breaking occurs in the simple-wave flow when all physical variables of the system can be
expressed as functions of only one of them what means that after transformation to the
corresponding Riemann invariants only one of them breaks and the others remain constant.
Gurevich and Meshcherkin claimed that this statement is correct also after formation of
a DSW, so that flows at both its edges have the same values of the non-breaking
Riemann invariants. This property is evidently correct in a simple case of self-similar
solutions of Whitham equations for the KdV equation studied in Ref.~\cite{gp-73}, when
Whitham equations are transformed to the diagonal form, and Gurevich and Meshcherkin
generalize it to situations when Riemann invariants of the modulation equations are
unknown or even do not exist.

A remarkable contribution into the general Whitham theory of modulations was made by
El~\cite{el-05} who showed that in simple-wave DSWs of Gurevich-Meshcherkin type it is
possible to find parameters of the DSW at its harmonic edge without full integration of the
Whitham modulation equations by using instead the Whitham
{\em conservation of number of waves} equation
\begin{equation}\label{eq1}
  \frac{\prt k}{\prt t}+\frac{\prt \om(k)}{\prt x}=0,
\end{equation}
where $k$ and $\om$ are the wave vector and
the frequency of a linear harmonic wave $\propto\exp[i(kx-\om t)]$ propagating along a
smooth background. El noticed that in the simple-wave breaking situation the physical
parameters at the small-amplitude DSW edge depend on a single parameter only.
As a result, Eq.~(\ref{eq1})
reduces to an ordinary differential equation which has an integral and the value of this
integral can be found with the use of the Gurevich-Meshcherkin conjecture. This yields
the value of the wave number of the wave packet at the small-amplitude edge of the DSW
which allows one to calculate the speed of this edge equal to the group velocity of the wave
packet provided the other parameters of the wave at this edge are also known.

The importance of the conservation of number of waves law (\ref{eq1}) was underlined by
Whitham \cite{whitham-65,whitham-74} who noticed that it is a direct
consequence of definitions of the wave vector $k=\prt\theta/\prt x$ and the frequency
$\om=-\prt\theta/\prt t$ in a slowly modulated wave, where $\theta=\theta(x,t)$ is the
phase of the wave. Therefore $k$ and $\om$ are defined also for nonlinear waves and
Eq.~(\ref{eq1}) fulfills along the entire DSW. As a result, it can be derived from any
full system of modulation equations. Unfortunately, Eq.~(\ref{eq1}) loses its meaning
at the soliton edge of a DSW where $k\to0$. In spite of that, El showed \cite{el-05} that
under some additional assumptions one can obtain from Eq.~(\ref{eq1}) the ordinary differential
equation relating the physical variables along the characteristic of Whitham equations
at the soliton edge of the DSW. Then one can get again the integral of this equation and find
its value by the same Gurevich-Meshcherkin method. This procedure gives the inverse
half-width $\tk$ of the leading soliton and $\tk$ is related to the soliton's velocity by
the well-known formula following from a simple reasoning: since a soliton propagates with
the same velocity as its tails and if the tails of a soliton have an exponential form
$\propto\exp[\mp\tk(x-V_s t)]$, $x\to\pm\infty$, then the tails obey the same linearized
equations which lead to the linear dispersion law $\om=\om(k)$. Hence, we arrive
immediately at the statement that the soliton
velocity $V_s$ is related with its inverse half-width $\tk$ by the formula
\begin{equation}\label{eq2}
  V_s={\tom(\tk)}/{\tk},
\end{equation}
where $\tom(\tk)$ is defined as
\begin{equation}\label{eq3}
  \tom(\tk)\equiv-i\om(i\tk).
\end{equation}
This relationship was noticed long ago and helped a lot in finding soliton solutions of
complicated systems of nonlinear wave equations (see, e.g., Refs.~\cite{ai-77,dkn-03}).

In an important particular case of initial step-like discontinuities all the parameters
besides $k$ and $\tk$ at the DSW edges are fixed, so after finding $k$ and $\tk$ by the
El method one can calculate such characteristics of the DSW as speeds of its edges and
the leading soliton amplitude. Application of this scheme to
the problems of evolution of initial step-like discontinuities for integrable equations,
when the global solutions for the whole DSWs are known due to existence of Riemann
invariants, showed that El's method works perfectly well at least for this class of
problems. Its further application to similar problems for non-integrable equations
and comparison of the results with numerical simulations demonstrated its good applicability
at least for moderate values of jumps of physical variables at the initial discontinuities.
As a result, a number of interesting problems was successfully considered by this method
(see, e.g., Refs.~\cite{egs-06,egkkk-07,egs-09,ep-11,hoefer-14,ckp-16,hek-17,ams-18}).

Obviously, `number of waves' conservation equation (\ref{eq1}) is universally correct for any
DSWs beyond those generated from step-like initial discontinuities and indeed it was
successfully used in estimate of asymptotic number of solitons generated from
an initially localized pulse evolved according to non-integrable equation \cite{egs-08}
(see also \cite{egkkk-07}). However, El's method cannot be applied directly
to the problems of evolution of pulses different from step-like discontinuities and
therefore it needs further development.

The aim of this paper is to develop the method of calculation of the main characteristics
of DSWs generated in evolution of more general pulses than the initial step-like discontinuities.
To this end, we study first whether Eq.~(\ref{eq1}) admits the transformation (\ref{eq3}), so that
the equation
\begin{equation}\label{eq4}
  \frac{\prt \tk}{\prt t}+\frac{\prt \tom(\tk)}{\prt x}=0
\end{equation}
plays at the soliton edge the role analogous to that of Eq.~(\ref{eq1}).
It is evident that, on the contrary to the situation with
modulated nonlinear periodic wave, where wavelength has clear enough physical meaning
along the DSW and $k$ can be defined as a gradient of the phase $\theta$, we cannot define
in a similar way the inverse `wave width' $\tk$, so Eq.~(\ref{eq4}) has clear sense at
the soliton edge only and even in this limit its correctness is not guaranteed.
In spite of that, a simple calculation shows that it is correct in the case of integrable
equations at the edge matching to the smooth profile of the pulse provided the initial
profile belongs to the class of simple waves which means physically that the pulse
propagates through medium `at rest' with constant waves of the dispersionless Riemann
invariants. Following to Gurevich and Meshcherkin \cite{gm-84}, we generalize this observation
to non-integrable equations. As a result, the conditions of applicability of Eq.~(\ref{eq4})
are formulated in terms of dispersionless wave breaking patterns and mean physically
that the wave breaking occurs at the boundary with the medium at rest. Naturally, these
conditions are fulfilled for the step-like discontinuities, hence El's theory is included
into this approach.

Assuming correctness
of Eq.~(\ref{eq4}), we can transform it again in vicinity of the soliton edge of the
DSW to the ordinary differential equation in simple-wave type problems, and then this
equation can be solved if an appropriate boundary condition at the small-amplitude edge is
known in the problem under consideration.
To find the law of motion of DSW's edge at the boundary with the smooth dispersionless
flow, we use the fact that in this limit the characteristic velocities of the Whitham
equations have known values what allows one to write down the corresponding limiting
Whitham equation in the hodograph transformed form and to solve it with the use of the fact
that the smooth dispersionless solution is also known. This procedure generalizes
to non-integrable equations the method used in Ref.~\cite{gkm-89} for calculation of
the law of motion of the small-amplitude edge in the KdV equation theory and applied recently
in Ref.~\cite{kamch-18} to calculation of the motion of the soliton edge at the boundary with
the smooth solution in the nonlinear Schr\"{o}dinger (NLS) equation theory.

Although in the case of monotonous initial pulses this method gives velocity of the edge
at the boundary with smooth non-uniform distribution only, in the case of non-monotonous pulse
the method provides also the asymptotic value of velocity of the opposite edge
at the boundary with the medium at rest. Comparison of the results obtained by this method with
known solutions of integrable KdV and NLS equations as well as with the results of numerical
solution of non-integrable equations confirms its validity. Thus, the method
greatly increases the area of applicability of the Whitham theory to description of
evolution of DSWs in nonlinear wave systems. In particular, it can be applied to
description of DSWs observed in experiments on evolution of pulses in shallow water
waves \cite{hs-78,tkco-16}, nonlinear optics \cite{wkf-07,Fat14,Xu17}, Bose-Einstein
condensates \cite{hoefer-06,ea-07,chang-08}.

The paper has the following structure. In Sec.~\ref{sec2} we discuss applicability of
Eqs.~(\ref{eq1}) and (\ref{eq4}) to the edges of DSWs for integrable KdV and NLS equations
and formulate the conditions of applicability of Eq.~(\ref{eq4}) to non-integrable
equations. The method formulated here is applied to different types of nonlinear wave
equations in Sec.~\ref{sec3} and situations with uni-directional and two-directional
propagation are considered separately in Secs.~\ref{sec3a} and \ref{sec3b}, respectively.
In both cases, we show that our method reproduces correctly the known results obtained
earlier for integrable KdV and NLS equations and then apply the method to typical in
nonlinear physics generalized KdV and NLS equations. In Sec.~\ref{sec4} the method
is validated by comparison of analytical formulas obtained in Whitham approximation with
exact numerical solution of the generalized NLS equation. Section \ref{sec5} is devoted
to conclusion and general discussion. Derivation of convenient for us form of El's equations
is given in the Appendix.

\section{Harmonic and soliton dispersion laws}\label{sec2}

We shall call the functions $\om=\om(k)$ and $\tom=\tom(\tk)$, which appear in Eqs.~(\ref{eq1})
and (\ref{eq4}), as ``harmonic'' and ``soliton'' dispersion laws, respectively. As was indicated
in the Introduction, these functions are defined by linearized equations of motion for small-amplitude
(harmonic) and soliton's tails limits, correspondingly, and they can be converted one into
another by means of Eq.~(\ref{eq3}). To formulate the conditions of applicability of Eqs.~(\ref{eq1})
and (\ref{eq4}) along DSW, we have to define $k$ and $\tk$ in the vicinity of DSW edges which can be done
if the periodic and soliton solutions of the equation under consideration are known. Moreover,
if these solutions are
parameterized by the Riemann invariants of the corresponding Whitham modulation equations, then
we can check the validity of Eqs.~(\ref{eq1}) and (\ref{eq4}) in the Whitham approximation. We
shall make such a check at the small-amplitude and soliton edges of DSWs described by the KdV and NLS
equations, and the results obtained will permit us to formulate a plausible conjecture
about the conditions of applicability of Eq.~(\ref{eq4}).

\subsection{KdV equation case and generalizations}

As is well known (see, e.g., \cite{kamch2000}), the KdV equation
\begin{equation}\label{eq3.122}
  u_t+6uu_x+u_{xxx}=0
\end{equation}
has a periodic solution which can be written in the form
\begin{equation}\label{eq3.148}
  u(x,t)=r_2+r_3-r_1-2(r_2-r_1)\sn^2(\sqrt{r_3-r_1}\,(x-Vt),m),
\end{equation}
where
\begin{equation}\label{eq7}
  V=2(r_1+r_2+r_3),\qquad m=\frac{r_2-r_1}{r_3-r_1}
\end{equation}
and $\sn$ is the Jacobi elliptic function. In the strictly periodic case the parameters
$r_1\leq r_2\leq r_3$ are constant, but in slowly modulated waves they become slow
functions of $x$ and $t$ whose evolution is governed by the Whitham equations
\begin{equation}\label{eq3.143}
  \frac{\prt r_i}{\prt t}+v_i(r_1,r_2,r_3)\frac{\prt r_i}{\prt x}=0,\quad i=1,2,3,
\end{equation}
where the expressions for the characteristic velocities $v_i(r)$ were obtained by
Whitham \cite{whitham-65,whitham-74} (see also \cite{kamch2000}).

In the small-amplitude limit $r_2-r_1\ll|r_2|$ the solution (\ref{eq3.148}) transforms
into a harmonic wave
\begin{equation}\label{eq6.14''}
\begin{split}
  & u(x,t)=r_3+(r_2-r_1)\cos[2\sqrt{r_3-r_1}\,(x-Vt)],\\
  & V=2(2r_1+r_3),
  \end{split}
\end{equation}
and in this limit the Whitham velocities are equal to
\begin{equation}\label{eq6.16}
  v_1|_{r_2=r_1}=v_2|_{r_2=r_1}=12r_1-6r_3, \quad v_3|_{r_2=r_1}=6r_3.
\end{equation}
The harmonic dispersion law follows immediately from linearized Eq.~(\ref{eq3.122}),
\begin{equation}\label{eq10}
  \om(k)=6uk-k^3,
\end{equation}
and it agrees with the expressions for $k$ and $\om$ that follow from (\ref{eq6.14''}),
\begin{equation}\label{eq11}
\begin{split}
  & k=2\sqrt{r_3-r_1},\\
  & \om=kV=4(2r_1+r_3)\sqrt{r_3-r_1},
  \end{split}
\end{equation}
if one takes into account that $u\approx r_3$ at this edge. Supposing that $r_1$ and $r_3$
evolve here according to the Whitham equations
\begin{equation}\label{eq12}
   \frac{\prt r_1}{\prt t}+(12r_1-6r_3)\frac{\prt r_1}{\prt x}=0,\qquad
   \frac{\prt r_3}{\prt t}+6r_3\frac{\prt r_3}{\prt x}=0,
\end{equation}
we readily find that the $k$ and $\om$ defined by Eqs.~(\ref{eq11}) satisfy Eq.~(\ref{eq1})
identically. This agrees with the general statement that the number of waves conservation
law (\ref{eq1}) is a strict consequence of the Whitham modulation equations.

\begin{figure}[t]
\centerline{\includegraphics[width=8cm]{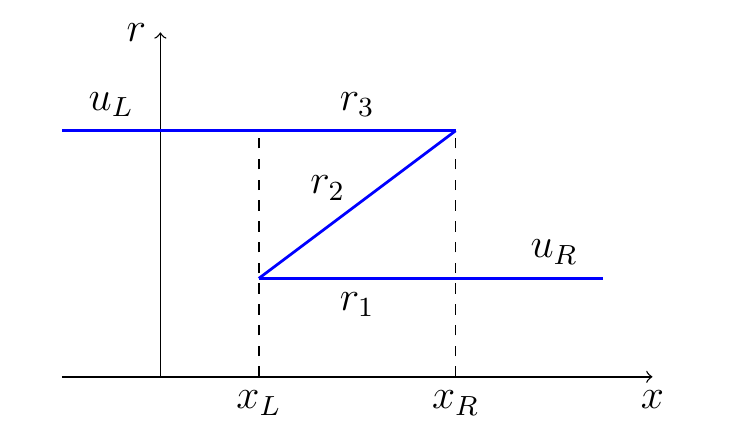}}
\caption{Plots of Riemann invariants for the solution of the step-like initial problem.
In this case $r_3=\mathrm{const}$ and Eq.~(\ref{eq4}) is correct.
}
\label{fig1}
\end{figure}

In a similar way, in the soliton limit $r_2\to r_3$ Eq.~(\ref{eq3.148}) reduces to
the soliton solution
\begin{equation}\label{eq6.14'}
\begin{split}
  & u(x,t)=r_1+\frac{2(r_3-r_1)}{\cosh^2[\sqrt{r_3-r_1}(x-V_st)]}, \\
  & V_s=2(r_1+2r_3),
  \end{split}
\end{equation}
and the Whitham velocities are given by the formulas
\begin{equation}\label{eq6.17}
  v_1|_{r_2=r_3}=6r_1, \quad v_2|_{r_2=r_3}=v_3|_{r_2=r_3}=2r_1+4r_3.
\end{equation}
Now the soliton dispersion law reads
\begin{equation}\label{eq14}
  \tom(\tk)=6u\tk+\tk^3
\end{equation}
and from Eq.~(\ref{eq6.14'}) we get $u\approx r_1$ and
\begin{equation}\label{eq15}
\begin{split}
  & \tk=2\sqrt{r_3-r_1},\\
  & \tom=\tk V_s=4(r_1+2r_3)\sqrt{r_3-r_1}.
  \end{split}
\end{equation}
Substitution of these expression into Eq.~(\ref{eq4}) with account of the Whitham
equations
\begin{equation}\label{eq16}
   \frac{\prt r_1}{\prt t}+6r_1\frac{\prt r_1}{\prt x}=0,\qquad
   \frac{\prt r_3}{\prt t}+(2r_1+4r_3)\frac{\prt r_3}{\prt x}=0
\end{equation}
yields after simple transformations the equation
\begin{equation}\label{eq17}
  \frac{\prt \tk}{\prt t}+\frac{\prt \tom(\tk)}{\prt x}=4\tk\frac{\prt r_3}{\prt x}.
\end{equation}
Hence, Eq.~(\ref{eq4}) does not hold for DSWs with changing Riemann invariant $r_3$
in the vicinity of the soliton edge. However, as is known from Gurevich-Pitaevskii theory \cite{gp-73} 
for evolution of the initial step-like discontinuity, in this particular case $r_3=\mathrm{const}$
and for such a DSW Eq.~(\ref{eq4}) is fulfilled. This type of DSWs corresponds to the
diagram of Riemann invariants shown in Fig.~1. Schematically, this diagram is
equivalent to a formal multi-valued solution of the Hopf equation,
\begin{equation}\label{eq18}
  u_t+6uu_x=0,
\end{equation}
which is the dispersionless approximation of the KdV equation, for the problem with
the initial step-like discontinuity, where $r_1,r_2,r_3$ symbolize the three values of
this multi-valued solution. (To avoid any confusion, we stress that numerically the
parameter $u$ in the formal dispersionless solution differs from values of $r_2$
obtained in the solution of the Whitham equations, and the same difference exists between
the coordinates $x_L,\,x_R$ of the edges in these two situations, that is we mean here
just qualitative geometric similarity of the respective diagrams rather than their
quantitative numerical identity.) It is natural to suppose that Eq.~(\ref{eq4}) remains
correct for DSWs described by non-integrable wave equations for some variable $u$,
if the corresponding dispersionless wave breaking pattern is given geometrically by
Fig.~\ref{fig1}, although its counterpart for the Riemann invariants of Whitham equations
does not exist anymore.
Actually, this conjecture corresponds exactly to the El theory \cite{el-05} applied
to the step-like initial problems.

\begin{figure}[t]
\centerline{\includegraphics[width=8cm]{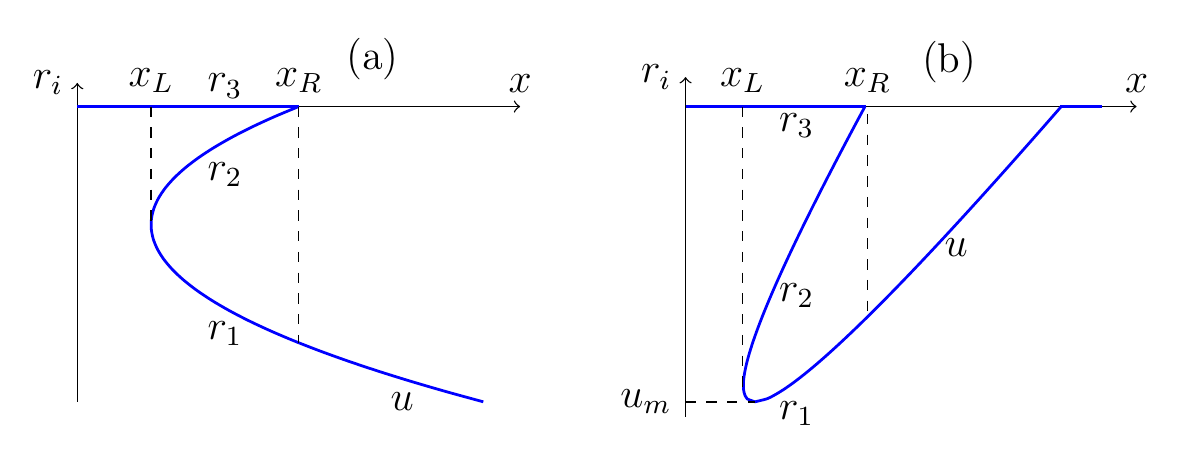}}
\caption{Plots of Riemann invariants for wave breaking of ``negative'' pulses
with monotonous (a) and localized (b) profiles.
In both cases $r_3=\mathrm{const}$ and Eq.~(\ref{eq4}) is correct.
}
\label{fig2}
\end{figure}

Now we notice that there exist other situations shown in Fig.~\ref{fig2} where $r_3=\mathrm{const}$
in the integrable KdV equation theory, whereas both $r_1$ and $r_2$ change with time and
space coordinate. The corresponding solutions of the Whitham equations were called
{\it quasi-simple} in Ref.~\cite{gkm-89}. We distinguish here the cases with monotonous and
non-monotonous initial pulses to underline that a non-monotonous initial pulse $u=u_0(x)$
is characterized also by some minimal value $u_m$ which plays an important role at the
asymptotic stage of evolution of the DSW evolved from such a pulse. In this type of DSWs,
the soliton edge $x_R$ propagates along a non-uniform and time-dependent background
evolving from $u_0(x)$, and the law of motion of this edge can be found with the use of
Eq.~(\ref{eq4}) bypassing the global solution of the full system of Whitham equations.
Again it is natural to assume that this method can be extended to non-integrable equations,
if the wave breaking pattern for a single wave variable $u$ has the same geometric form
and the dispersion and nonlinearity are such that the soliton edge is located at the
boundary with the dispersionless smooth solution.

\begin{figure}[t]
\centerline{\includegraphics[width=8cm]{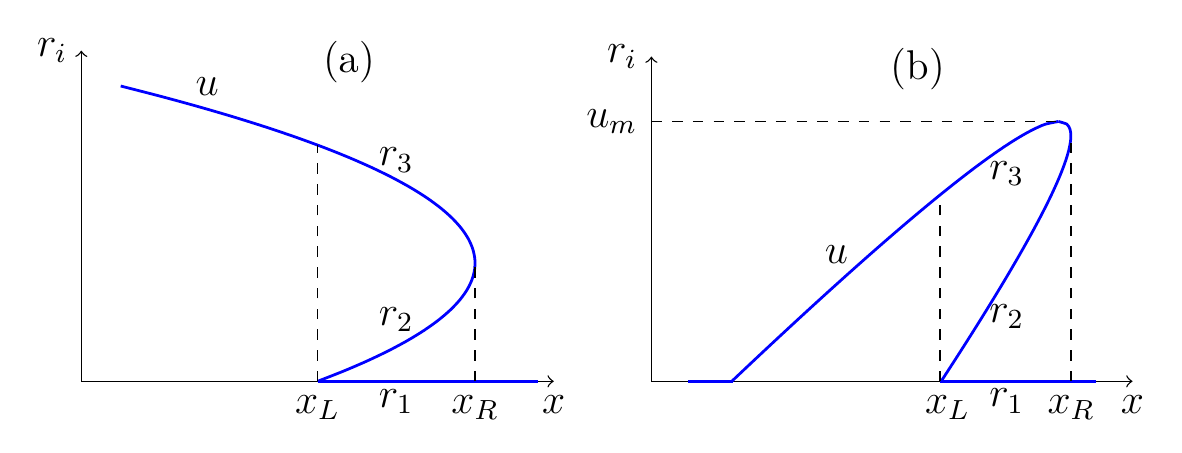}}
\caption{Plots of Riemann invariants for wave breaking of ``positive'' pulses
with monotonous (a) and localized (b) profiles.
In both cases $r_3$ changes with $x$ and Eq.~(\ref{eq4}) is not applied.
}
\label{fig3}
\end{figure}

On the contrary, if the wave breaking pattern
has the form depicted in Fig.~\ref{fig3}, then $r_3$ is not constant, Eq.~(\ref{eq4}) loses
its applicability and the law of motion of the soliton edge $x_R$ cannot be found by
solving this equation. It is known that such a localized positive pulse evolves asymptotically
into a sequence of well-separated solitons and in the KdV equation case the velocity of the leading
soliton can be found with the use of the Karpman formula \cite{karpman-68}. The number of solitons
generated from the initial pulse can be calculated in the non-integrable case by the method of
Refs.~\cite{egkkk-07,egs-08}.
The universal Eq.~(\ref{eq1}) is always correct and, as
we shall see, permits one to find the law of motion of the small-amplitude edge $x_L$.
To distinguish these two situations, we shall call them ``negative'' in the case of Fig.~\ref{fig2}
and ``positive'' in the case of Fig.~\ref{fig3}. It is important for us that for a proper relative
sign of nonlinear and dispersive effects, in the first case the soliton edge is located at
the boundary $x_R$ with the smooth solution $u$ and in the second case the small-amplitude
edge is located at such a boundary $x_L$.

At the same time we should notice that although Eq.~(\ref{eq1}) is always applicable,
in situations shown in Fig.~\ref{fig2} the small-amplitude
edge propagates into the region with constant $u$ which is a trivial solution of the
Hopf equation (\ref{eq18}), that is there is no explicit relationship between the values of $u_L$,
$x_L$ and $t$ at the small-amplitude edge. Therefore the law of motion of this edge
cannot be found by the method suggested here. In spite of that, in the case of an initially
localized pulse depicted in Fig.~\ref{fig2}(b),
the asymptotic value of the velocity of the small-amplitude edge can be expressed in
terms of the minimal value $u_m$ of the initial distribution. All these problems will be
considered in Sec.~\ref{sec3}.

\subsection{NLS equation case and generalizations}

Now we turn to the question of validity of Eq.~(\ref{eq4}) for DSWs whose evolution is
described by the NLS equation
\begin{equation}\label{eq19}
    i\psi_t+\frac12\psi_{xx}-|\psi|^2\psi=0.
\end{equation}
In this case it is convenient to represent the periodic solution in terms of the variables
$\rho$ and $u$ such that
\begin{equation}\label{eq20}
    \psi(x,t)=\sqrt{\rho(x,t)}\exp\left(i\int^xu(x',t)dx'\right).
\end{equation}
Then, for example, in the physical context of Bose-Einstein condensates, $\rho$ has a meaning
of its density and $u$ of its flow velocity. In terms of these variables, the NLS equation
can be written as a system of hydrodynamics-like equations
\begin{equation}\label{eq20b}
    \begin{split}
    & \rho_t+(\rho u)_{x}=0,\\
    & u_t+uu_{x}+\rho_{x}+\left(\frac{\rho_{x}^2}{8\rho^2}
    -\frac{\rho_{xx}}{4\rho}\right)_{x}=0,
    \end{split}
\end{equation}
where the last term in the second equations describes the dispersion effects. In smooth flows,
when higher derivatives are small, this term can be neglected and we arrive at the dispersionless
limit represented by the `shallow water equations'
\begin{equation}\label{eq20c}
        \rho_t+(\rho u)_{x}=0,\quad
    u_t+uu_{x}+\rho_{x}=0.
\end{equation}
The periodic solution of the system (\ref{eq20b}) is given by the formulas
(see, e.g., \cite{kamch2000})
\begin{equation}\label{eq21}
\begin{split}
     \rho(x,t)=&\tfrac14(r_4-r_3-r_2+r_1)^2+(r_4-r_3)(r_2-r_1)\\
    &\times\mathrm{sn}^2(\sqrt{(r_4-r_2)(r_3-r_1)}(x-Vt),m),\\
     u(x,t)=&V+\frac{j}{\rho(x,t)},
    \end{split}
\end{equation}
where
\begin{equation}\label{3-13}
\begin{split}
     V=&\frac12(r_1+r_2+r_3+r_4),\quad
    m=\frac{(r_2-r_1)(r_4-r_3)}{(r_4-r_2)(r_3-r_1)},\\
     j=&\frac18(-r_1-r_2+r_3+r_4)(-r_1+r_2-r_3+r_4)\\
    &\times(r_1-r_2-r_3+r_4),
    \end{split}
\end{equation}
that is $j$ has the meaning of the density current in the reference frame where the phase
velocity $V$ is equal to zero. The parameters $r_1\leq r_2\leq r_3\leq r_4$ play the role
of the Riemann invariants in a modulated wave so that the Whitham equations have a
diagonal form
\begin{equation}\label{eq23}
  \frac{\prt r_i}{\prt t}+v_i(r)\frac{\prt r_i}{\prt x}=0,\quad i=1,2,3,4,
\end{equation}
and expressions for the velocities $v_i$ were found in Refs.~\cite{FL-87,pavlov-87}.
At the small-amplitude edge $r_3=r_4$ we have a harmonic wave
\begin{equation}\label{eq26}
\begin{split}
    \rho=&\frac14(r_2-r_1)^2-\frac12(r_2-r_1)(r_4-r_3)\\
    &\times\cos\left[2\sqrt{(r_4-r_1)(r_4-r_2)}
    (x-Vt)\right],\\
    V=&\frac12(r_1+r_2+2r_4),
    \end{split}
\end{equation}
and the velocities $v_i$ are given here by
\begin{equation}\label{eq25}
\begin{split}
    & v_1=\frac12(3r_1+r_2),\quad v_2=\frac12(r_1+3r_2),\\
    & v_3=v_4=2r_4-\frac{(r_2-r_1)^2}{2(2r_4-r_1-r_2)}.
    \end{split}
\end{equation}
(Analogous formulas can be obtained in another small-amplitude limit $r_2=r_1$, but we
shall not need them in what follows.) At the soliton edge we have the dark soliton solution
\begin{equation}\label{eq27}
\begin{split}
&\rho=\frac{1}{4}(r_4 - r_1)^2 -
\frac{(r_4 - r_2)(r_2 - r_1)}{\cosh^2 [\sqrt{(r_4 - r_2)(r_2 - r_1)}(x-V_st)]},\\
&V_s=\frac{1}{2}(r_1+2r_2+r_4) ,
\end{split}
\end{equation}
and near the soliton edge the characteristic velocities of the Whitham system are given by
the formulas
\begin{equation}\label{eq24}
    \begin{split}
    & v_1=\frac12(3r_1+r_4),\quad v_2=v_3=\frac12(r_1+2r_2+r_4),\\
    & v_4=\frac12(r_1+3r_4).
    \end{split}
\end{equation}
Now we can check the validity of Eqs.~(\ref{eq1}) and (\ref{eq4}) in framework of the Whitham
theory for the NLS equation.

At the small-amplitude edge Eq.~(\ref{eq26}) gives the expressions for the wave vector and the frequency,
\begin{equation}\label{eq28}
\begin{split}
  & k=2\sqrt{(r_4-r_1)(r_4-r_2)},\\
  & \om=kV=(r_1+r_2+2r_4)\sqrt{(r_4-r_1)(r_4-r_2)}.
   \end{split}
\end{equation}
Their substitution into Eq.~(\ref{eq1}) with account of the Whitham equations (\ref{eq23}) with velocities
(\ref{eq25}) demonstrate after simple calculation that Eq.~(\ref{eq1}) is fulfilled identically.
[Similar calculation proves the validity of Eq.~(\ref{eq1}) in another small-amplitude limit $r_2=r_1$.]
Thus, we have confirmed again that the number of waves conservation law is a consequence of the
Whitham equations in agreement with the general statement of Whitham \cite{whitham-65,whitham-74}.

At the soliton edge we find from Eq.~(\ref{eq27}) that
\begin{equation}\label{eq29}
\begin{split}
  & \tk=2\sqrt{(r_4-r_1)(r_4-r_2)},\\
  & \tom=(r_1+2r_2+r_4)\sqrt{(r_4-r_1)(r_4-r_2)},
  \end{split}
\end{equation}
and substitution of these expressions into Eq.~(\ref{eq1}) with account of Whitham equations
(\ref{eq23}) with velocities (\ref{eq24}) gives
\begin{equation}\label{eq30}
  \frac{\prt \tk}{\prt t}+\frac{\prt \tom(\tk)}{\prt x}=2\tk\frac{\prt r_2}{\prt x}.
\end{equation}
Here we should make an important remark. In the above expressions for the soliton limit
we denoted the common value of the two Riemann invariants $r_3=r_2$ as $r_2$ which explains
the appearance of $r_2$ in the right-hand side of Eq.~(\ref{eq30}). If we had denoted it
as $r_3$, then we would have obtained Eq.~(\ref{eq30}) with $r_2$ replaced by $r_3$.
This restores the two-directional symmetry of the NLS equation.

\begin{figure}[t]
\centerline{\includegraphics[width=7.5cm]{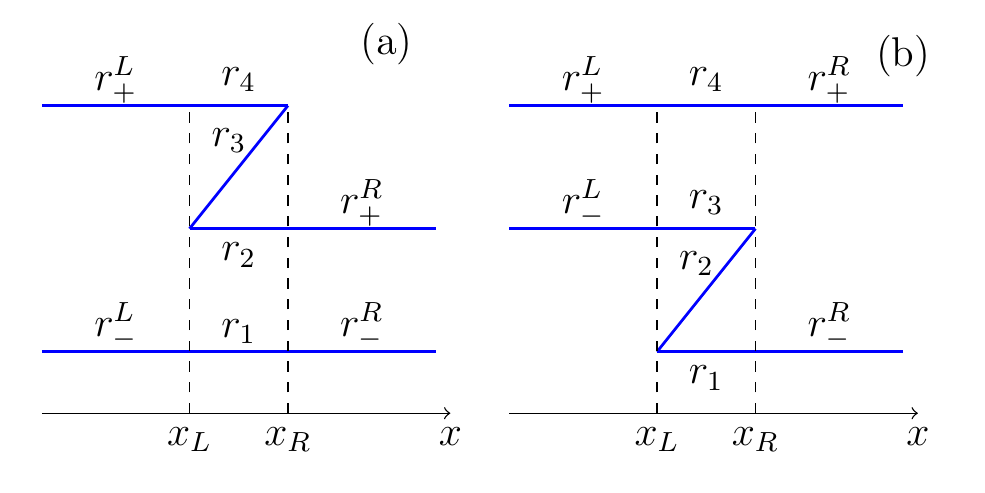}}
\caption{Diagrams of Riemann invariants for wave breaking of step-like pulses
in NLS equation theory.
}
\label{fig4}
\end{figure}

Thus, we arrive at the conclusion that Eq.~(\ref{eq1}) is universally correct and (\ref{eq4})
is correct when in the corresponding diagram of Riemann invariants we have either $r_2$ or
$r_3$ constant. Obviously, this takes place, for example, in the diagrams shown in Fig.~\ref{fig4},
where $r_i$ ($i=1,2,3,4$) denote the Riemann invariants of of the Whitham equations and
$r_{\pm}^L$, $r_{\pm}^R$ denote the Riemann invariants
\begin{equation}\label{eq31}
  r_{\pm}=\frac{u}2\pm\sqrt{\rho}
\end{equation}
of the dispersionless system (\ref{eq20c}) which takes a diagonal Riemann form in terms of
these variables,
\begin{equation}\label{eq32}
\begin{split}
   & \frac{\prt r_+}{\prt t}+v_+(r_+,r_-)\frac{\prt r_+}{\prt x}=0,\\
   & \frac{\prt r_-}{\prt t}+v_-(r_+,r_-)\frac{\prt r_-}{\prt x}=0,
    \end{split}
\end{equation}
where
\begin{equation}\label{3-8}
\begin{split}
    & v_+(r_+,r_-)=\frac12(3r_++r_-),\\
    &  v_-(r_+,r_-)=\frac12(r_++3r_-).
    \end{split}
\end{equation}
These velocities coincide exactly with the corresponding limits of the Whitham velocities
which provides continuous matching of the Riemann invariants of Whitham and dispersionless
equations as is shown in Fig.~\ref{fig4}.

The diagrams shown in Fig.~\ref{fig4} symbolize the step-like wave breaking of simple waves,
so that in Fig.~\ref{fig4}(a) the invariant $r_+$ of the right-propagating wave breaks and $r_-$
is constant whereas in Fig.~\ref{fig4}(b) the invariant $r_-$ of the left-propagating wave
breaks and $r_+$ remains constant. The Gurevich-Mescherkin conjecture claims that
even if the Riemann invariants of the Whitham modulation equations do not exist,
nevertheless the values of the dispersionless Riemann invariant $r_-$ is transferred
through the DSW generated after wave breaking of the right-propagating simple wave and,
in a similar way, the value of $r_+$ is transferred through the DSW generated after the
wave breaking of the left-propagating wave. El's theory \cite{el-05} corresponds to
this situation and provides the method of calculation of parameters of the edges of
DSWs generated from initial step-like discontinuities for the non-integrable wave
equations case. The agreement of El's theory with numerical simulations shows that
although the invariants $r_2$ or $r_3$ do not exist along the whole DSW, their
existence at its edges ($r_2=r_+^R$ in Fig.~\ref{fig4}(a) or $r_3=r_-^L$ in Fig.~\ref{fig4}(b))
is enough for finding the velocities of DSW edges in the case of step-like initial
problems. Here we assume correctness of Eq.~(\ref{eq4}) in vicinity of the DSW edge. Actually,
this statement means generalization of the Gurevich-Mescherkin conjecture to quasi-simple
waves where either $r_{\pm}^L$ or $r_{\pm}^R$ are constant at one of the edges of the DSW.
As is clear from Eqs.~(\ref{eq31}), in the NLS equation case the constancy of both
dispersionless Riemann invariants at one of the edges of the DSW means that this DSW
propagates into a uniform ($\rho_0=\mathrm{const}$) region with constant flow velocity
($u_0=\mathrm{const}$) which can be put equal to zero in the appropriate reference frame.
We can formulate these observations as a conjecture that Eq.~(\ref{eq4}) holds for
DSWs propagating into uniform `quiescent' medium.

\begin{figure}[t]
\centerline{\includegraphics[width=8cm]{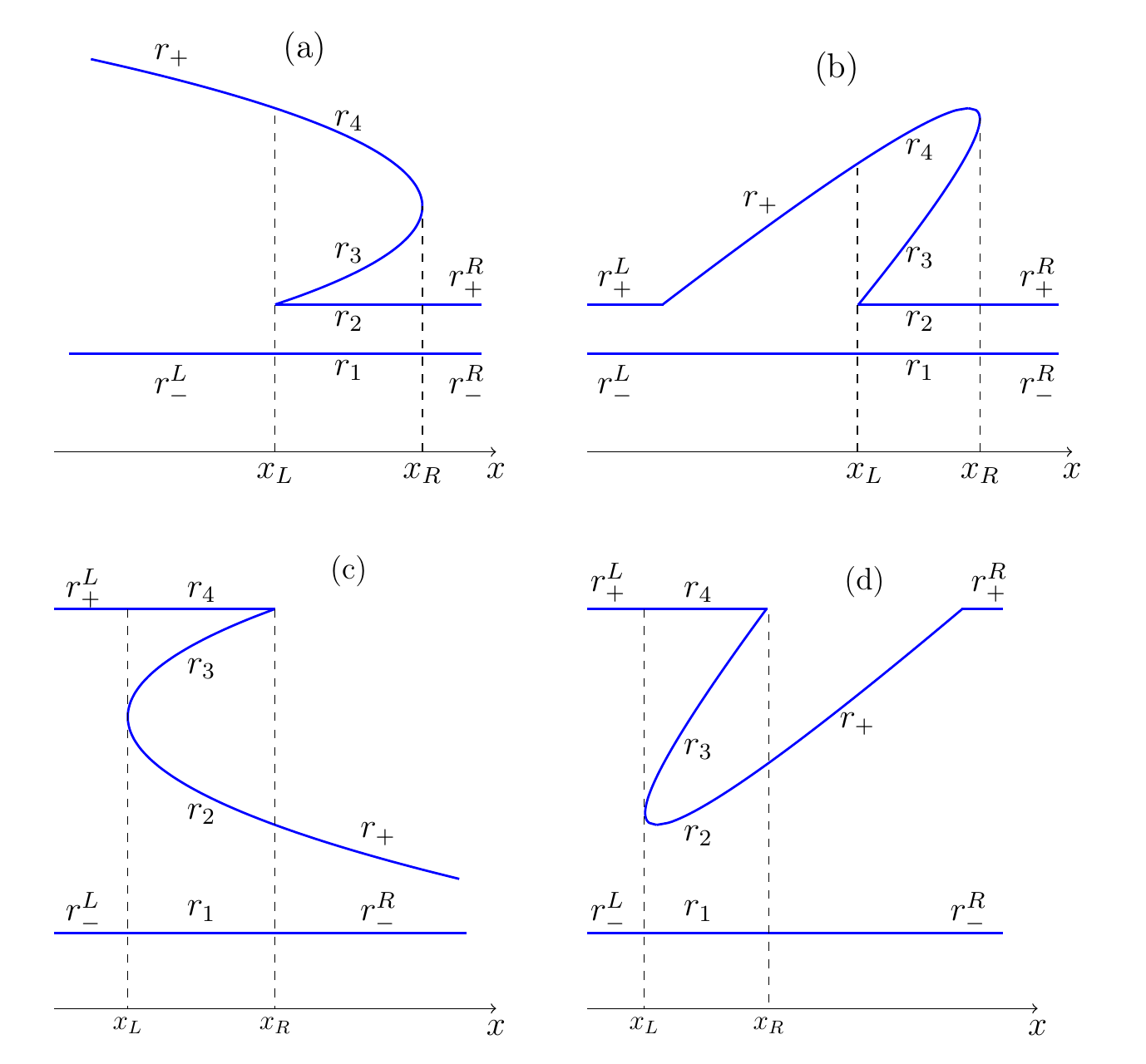}}
\caption{Diagrams of Riemann invariants for wave breaking of ``positive'' pulses
with monotonous (a) and localized (b) profiles of dispersionless Riemann invariant $r_+$.
In both cases $r_2=\mathrm{const}$ and Eq.~(\ref{eq4}) is correct. For ``negative''
initial distributions (c) and (d) of the invariant $r_+$ the invariants $r_2$ and $r_3$
of the Whitham modulation system vary with $x$ and Eq.~(\ref{eq4}) is not fulfilled.
}
\label{fig5}
\end{figure}

Now we notice that, as in the KdV equation case, there exist other situations in the NLS
equation theory where $r_2$ or $r_3$ are constant. In Fig.~\ref{fig5}(a,b) the Riemann invariant
$r_+$ with ``positive'' initial distribution breaks and we distinguish again here monotonous
and localized pulses; in Fig.~\ref{fig5}(c,d) we have depicted analogous situations
with breaking of $r_+$-invariant for ``negative'' initial profiles.
We have shown recently in Ref.~\cite{kamch-18} that in the case corresponding to
Fig.~\ref{fig5}(a) the law of motion of the soliton edge $x_L(t)$ can be found without
knowing the global solution of the Whitham equations. It is natural to suppose that
this method can be generalized in such a way that the law of motion of the soliton edge
can be found with the use of Eq.~(\ref{eq4}) for pulses whose evolution is governed by
non-integrable wave equations, if the wave breaking patterns of the dispersionless
Riemann invariants coincide geometrically with those shown in Fig.~\ref{fig5}(a,b) in spite of that
the Riemann invariants of the Whitham modulation equations do not exist anymore.
Due to universal applicability of Eq.~(\ref{eq1}), the laws of motion of the small-amplitude
edges at the boundaries with smooth solutions for breaking Riemann invariants can be found in
situations whose integrable counterparts are depicted in Figs.~\ref{fig5}(c,d) where $r_2$
and $r_3$ are not constant. In these cases, the small-amplitude edges $x_R$
propagate into smooth distributions of the dispersionless Riemann invariants $r_+$.
The laws of motion of the soliton edges $x_L$
can be found only at the asymptotic stage of evolution, when the initial pulse evolves into a
sequence of well separated solitons. In integrable cases this stage can be studied with the use
of the Bohr-Sommerfeld quantization rule for the associated Lax spectral problem (see, e.g.,
\cite{kku-02}) and in non-integrable cases the number of solitons can be calculated by the
method suggested in Refs.~\cite{egkkk-07,egs-08}.

Now, after formulation of the class of problems where the number of waves conservation law
(\ref{eq1}) and its soliton counterpart (\ref{eq4}) are applicable, we can proceed to
demonstration of their concrete applications to DSWs evolutions.

\section{Motion of dispersive shock edges}\label{sec3}

Evolution of DSWs generated from initial step-like discontinuities were studied in much detail in
Refs.~\cite{el-05,egs-06,egkkk-07,egs-09,ep-11,hoefer-14,ckp-16,hek-17,ams-18} by El's method
and we shall not consider this particular case here. Instead, we shall turn to the class of
problems referred to in Ref.~\cite{gkm-89} as {\it quasi-simple} DSWs. To illustrate the correctness
of our approach, we shall consider first integrable situations where our results can be compared
with the results known from solutions obtained by the inverse scattering transform method.

To simplify exposition, we notice here that El's method together with the Gurevich-Meshcherkin
conjecture can be formulated as an `extrapolation' procedure, that is, instead of speaking about
solving Eqs.~(\ref{eq1}) and (\ref{eq4}) along edge characteristics of the Whitham system with
the Gurevich-Meshcherkin initial condition, we say that we extrapolate solutions of the limiting
Whitham equation (\ref{eq1}) or (\ref{eq4}) from the vicinity of one edge to the whole DSW and
impose the proper boundary condition at the opposite edge. This formulation gives the same
results as El's method and has some advantages for generalizations of the method to
non-step-like initial conditions.

\subsection{Unidirectional propagation}\label{sec3a}

We suppose that our nonlinear wave is described by a single variable $u$. To illustrate the method, we
consider first the completely integrable KdV equation, but we shall use here only
the general properties of the Whitham modulation equations not related with their
explicit diagonal form.

\subsubsection{KdV equation: Positive pulse}

We shall start with a monotonous pulse with the initial form $u_0(x)>0$ shown in Fig.~\ref{fig3}(a).
Our task here is to find the law of motion
of the small-amplitude edge $x_L(t)$. In the region $x<x_L(t)$ the pulse is smooth and its
evolution is described by the dispersionless Hopf equation (\ref{eq18}) whose solution reads
\begin{equation}\label{eq34}
  x-6ut=\ox(u),
\end{equation}
where $\ox(u)$ is a function inverse to $u_0(x)$. On the other hand, the function $u(x)$ can be
treated at $x=x_L$ as a solution of the Whitham equations in the limit $x\to x_L+0$
corresponding to the characteristic velocity $v_+=6\overline{u}$ of the Whitham system,
where $\overline{u}$ is understood as a mean value of the wave variable at this
small-amplitude limit. But, according to the general principles of the Whitham theory
\cite{whitham-65,whitham-74}, the system of Whitham modulation equations has in this
limit another characteristic velocity $v_-$ equal to the group velocity
\begin{equation}\label{eq35}
  v_-=\frac{dx_L}{dt}=\frac{d\om}{dk}=6\overline{u}-3k^2
\end{equation}
of the small-amplitude wave at the edge $x_L$.
The wave vector $k$ and the frequency $\om$ change at this edge in such a way that Eq.~(\ref{eq1})
is fulfilled,
\begin{equation}\label{eq36}
  \frac{\prt k}{\prt t}+\frac{\prt \om}{\prt x}=0,\qquad \om(\overline{u},k)=6\overline{u}k-k^3.
\end{equation}
If we suppose that we deal here with a simple-wave type of solutions of the Whitham system,
which agrees with the form (\ref{eq34}) of the smooth solution, then $k$ depends on $x$ and $t$
via $\overline{u}(x,t)$, $k=k(\overline{u})$. Taking into account that at this edge
Eq.~(\ref{eq18}) holds, we reduce
Eq.~(\ref{eq36}) to an ordinary differential equation
\begin{equation}\label{eq37}
  k\frac{dk}{d\overline{u}}=2.
\end{equation}
Now we solve this equation and extrapolate the solution to the whole DSW with the boundary condition
that at the opposite soliton edge the distance between solitons tends to infinity,
\begin{equation}\label{eq38}
  k=0\qquad\text{at}\qquad \overline{u}=0.
\end{equation}
This extrapolation gives a correct value
\begin{equation}\label{eq39}
  k=2\sqrt{u}
\end{equation}
of the wave number at the edge $x_L(t)$ where $u=\ou$. Actually, this procedure is equivalent
to solving the El equation \cite{el-05}, but we prefer to use here directly the number of waves
conservation law Eq.~(\ref{eq36}) to make the method more transparent and flexible for further
generalizations.

Substitution of Eq.~(\ref{eq39}) into Eq.~(\ref{eq35}) gives the value of the characteristic
velocity
\begin{equation}\label{eq40}
  v_-=\frac{dx_L}{dt}=-6u,
\end{equation}
which corresponds to the limiting form of the Whitham equation
\begin{equation}\label{eq41}
  \frac{\prt x_L}{\prt u}+6u\frac{\prt t}{\prt u}=0
\end{equation}
written in the hodograph transform representation (see, e.g., \cite{kamch2000}).
This equation must be compatible with Eq.~(\ref{eq34}) taken at $x=x_L$, so that
elimination of $x_L$ yields the differential equation
\begin{equation}\label{eq42}
  2u\frac{dt}{du}+t=-\frac16\frac{d\ox}{du}.
\end{equation}
We suppose that wave breaking occurs at the moment $t=0$ at the boundary $u=0$ of the
initial pulse. Hence, Eq.~(\ref{eq42}) must be solved with the initial condition
\begin{equation}\label{eq43}
  t(0)=0
\end{equation}
which gives at once
\begin{equation}\label{eq44}
  t(u)=-\frac1{12\sqrt{u}}\int_0^u\frac{\ox'(u)}{\sqrt{u}}\,du
\end{equation}
and substitution of this expression into Eq.~(\ref{eq34}) yields
\begin{equation}\label{eq45}
  x_L(u)=-\frac{\sqrt{u}}2\int_0^u\frac{\ox'(u)}{\sqrt{u}}\,du+\ox(u).
\end{equation}
It is assumed here (and in similar situations in what follows) that the
function $\ox(u)$ vanishes in the limit $u\to0$ fast enough so that the
integrals converge and tend to zero in this limit to fulfill the initial
condition (\ref{eq43}).
The formulas (\ref{eq44}) and (\ref{eq45}) give us the law of motion of the
small-amplitude edge $x_L(t)$ in parametric form for a positive monotonous profile
$u_0(x)$ of the initial pulse. For example, in the case of a parabolic profile
$u_0(x)=\sqrt{-x}$, we have $\ox(u)=-u^2$,  and easy calculation reproduces the
well-known result \cite{gkm-89,ks-90} (see also \cite{kamch2000})
\begin{equation}\label{eq46}
  x_L(t)=-27t^2.
\end{equation}
Our approach can be considered as a modification of calculation presented in
Ref.~\cite{gkm-89}, but, instead of using the known expression for the characteristic
velocity $v_-$ in terms of Riemann invariants, we have calculated it in Eq.~(\ref{eq40})
by means of El's rule which is applicable equally to both integrable
and non-integrable wave equations.

\begin{figure}[t]
\centerline{\includegraphics[width=8cm]{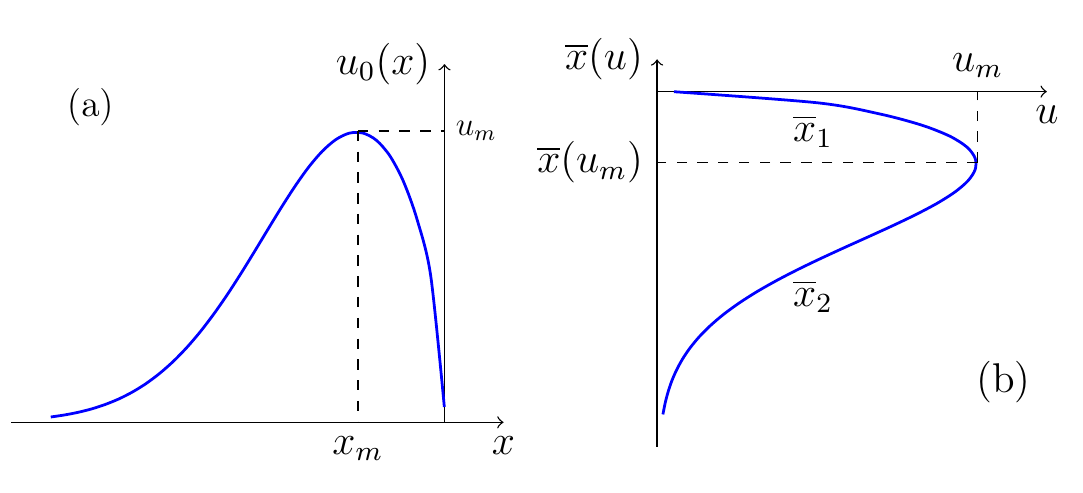}}
\caption{(a) Initial profile $u_0(x)$ of a localized pulse. (b) The inverse function
$\ox(u)$ is represented by
two branches $\ox_1(u)$ and $\ox_2(u)$.
}
\label{fig6}
\end{figure}

In the case of a localized initial pulse Fig.~\ref{fig6}(a), the inverse function becomes two-valued
and we denote its two branches as $\ox_1(u)$ and $\ox_2(u)$ (see Fig.~\ref{fig6}(b)). The formulas
(\ref{eq44}) and (\ref{eq45}) remain true up to the moment
\begin{equation}\label{eq47}
  t_m=-\frac1{12\sqrt{u_m}}\int_0^{u_m}\frac{\ox'_1(u)}{\sqrt{u}}\,du
\end{equation}
at which the small-amplitude edge reaches the coordinate
\begin{equation}\label{eq48}
  x_m=-\frac{\sqrt{u_m}}2\int_0^{u_m}\frac{\ox'_1(u)}{\sqrt{u}}\,du+\ox_1(u_m).
\end{equation}
After that the edge $x_L(t)$ propagates along the second branch $x-6ut=\ox_2(u)$ of
the dispersionless solution and the equation
\begin{equation}\label{eq50}
  2u\frac{dt}{du}+t=-\frac16\frac{d\ox_2}{du}.
\end{equation}
must be solved with the initial condition
\begin{equation}\label{eq51}
  t(u_m)=t_m
\end{equation}
which yields
\begin{equation}\label{eq52}
  t(u)=-\frac1{12\sqrt{u}}\left(\int_0^{u_m}\frac{\ox'_1(u)}{\sqrt{u}}\,du
  +\int_{u_m}^u\frac{\ox'_2(u)}{\sqrt{u}}\,du\right)
\end{equation}
and, hence,
\begin{equation}\label{eq53}
  x_L(u)=-\frac{\sqrt{u}}2\left(\int_0^{u_m}\frac{\ox'_1(u)}{\sqrt{u}}\,du
  +\int_{u_m}^u\frac{\ox'_2(u)}{\sqrt{u}}\,du\right)+\ox_2(u).
\end{equation}

The law of motion of the soliton edge cannot be found by this method because the
initial pulse Fig.~\ref{fig6}(a) does not correspond to the wave breaking pattern
(see Fig.~\ref{fig2}) to which the soliton dispersion equation (\ref{eq4}) is applicable.
For the integrable KdV equation case, this law of motion can be found by the
inverse scattering transform method (see Refs.~\cite{gkme-92,kud-92}).
As is known, in this case the initial pulse evolves to a sequence of solitons whose
number can be calculated with the use of the Karpman formula \cite{karpman-68}
which admits the ``non-integrable'' generalization \cite{egs-08,egkkk-07}.

Although the law of motion $x_R=x_R(t)$ remain unknown, its asymptotic behavior can be
found with the use of our extrapolation procedure. Let us consider the vicinity of the
moment of time $t=t_m$ when the small-amplitude edge of DSW reaches the point $x_m$
where $u$ takes its maximal value $u_m$. In terms of Riemann invariants this corresponds to
the maximal value of $r_3$, that is $\prt r_3/\prt x=0$, and Eq.~(\ref{eq17}) reduces to
Eq.~(\ref{eq4}). Then at vicinity of this point we have $\tk=\tk(\ou)$ and Eq.~(\ref{eq4})
transforms to the equation
\begin{equation}\label{eq53b}
  \tk\,\frac{d\tk}{d\ou}=-2
\end{equation}
which should be solved with the boundary condition $\tk\to0$ at $u\to u_m$, that is
the amplitude of solitons tends to zero together with $\tk$ at the small-amplitude
edge. Then the solution reads
\begin{equation}\label{eq53c}
  \tk=2\sqrt{u_m-u}
\end{equation}
and at the soliton edge with $u=0$ the inverse half-width of solitons reaches its
maximal value $\tk_m=2\sqrt{u_m}$ which corresponds to maximal soliton velocity
and its maximal amplitude,
\begin{equation}\label{eq53d}
  v_s={\tom}/{\tk}=\tk^2=4u_m, \qquad a=2u_m,
\end{equation}
which agrees with known results for the KdV equation.
Thus, at the asymptotic stage of evolution the leading soliton propagates according to the law
$x_R\approx 4u_mt$.

\subsubsection{KdV equation: Negative pulse}

Let us now have a negative monotonous initial pulse $u_0(x)<0$, $x>0$,
with the inverse function $x=\ox(u)$. Then the smooth solution
is given in the dispersionless limit by
\begin{equation}\label{eq54}
  x-6ut=\ox(u),\quad u<0,\quad x>0,
\end{equation}
and it breaks at the rear small amplitude edge. The soliton edge propagates along the
non-uniform background $u(x,t)$ represented by the solution (\ref{eq54}). In this case,
the wave breaking pattern has the form of Fig.~\ref{fig2}(a) and, according to our conjecture,
Eq.~(\ref{eq4}) is applicable,
\begin{equation}\label{eq55}
  \frac{\prt \tk}{\prt t}+\frac{\prt \tom(\tk)}{\prt x}=0,\quad \tom(\ou,\tk)=6\ou\tk+\tk^3.
\end{equation}
We suppose again that the soliton inverse width $\tk$ depends on $x$ and $t$
via $\ou(x,t)$, $\tk=\tk(\ou)$, and with account of the dispersionless equation
$\ou_t+6\ou\,\ou_x=0$ valid at the soliton edge we reduce Eq.~(\ref{eq55}) to the
ordinary differential equation
\begin{equation}\label{eq56}
  \tk\frac{d\tk}{d\ou}=-2.
\end{equation}
Extrapolation of this equation to the whole DSW with the boundary condition that the soliton
inverse width vanishes together with its amplitude [see Eq.~(\ref{eq6.14'})], that is $\tk(0)=0$,
give at once the value of $\tk$ at the soliton edge,
\begin{equation}\label{eq58}
  \tk=2\sqrt{-u}.
\end{equation}
Then the soliton edge velocity is equal to
\begin{equation}\label{eq59}
  \tom/\tk=6u+\tk^2=2u,
\end{equation}
and this gives us the characteristic velocity
\begin{equation}\label{eq60}
  v_+=\frac{dx_R}{dt}=2u
\end{equation}
of the Whitham equation
\begin{equation}\label{eq61}
  \frac{\prt x_R}{\prt u}-2u\frac{\prt t}{\prt u}=0
\end{equation}
written in hodograph form. The compatibility condition of Eqs.~(\ref{eq54}) and (\ref{eq61})
leads to the differential equation
\begin{equation}\label{eq62}
  2u\frac{dt}{du}+3t=-\frac12\ox'(u),
\end{equation}
whose solution with the initial condition $t(0)=0$ yields
\begin{equation}\label{eq63}
  t(u)=\frac1{4(-u)^{3/2}}\int_0^u\sqrt{-u}\,\ox'(u)du
\end{equation}
and, hence,
\begin{equation}\label{eq64}
  x_R=-\frac3{2\sqrt{-u}}\int_0^u\sqrt{-u}\,\ox'(u)du+\ox(u).
\end{equation}
These formulas give us the dependence $x_R(t)$ in parametric form. In particular,
in the case of the initial pulse with the form $u_0(x)=-x^{1/n}$, $\ox(u)=(-u)^n$, we obtain
\begin{equation}\label{eq65}
  x_R(t)=-2\left(1-\frac1n\right)\left(4+\frac2n\right)^{\frac1{n-1}}t^{\frac{n}{n-1}},
\end{equation}
and this result can be confirmed by the global solution of the Whitham equations
obtained by the methods based on the complete integrability of the KdV equation
\cite{kamch-18b}.

\begin{figure}[t]
\centerline{\includegraphics[width=8cm]{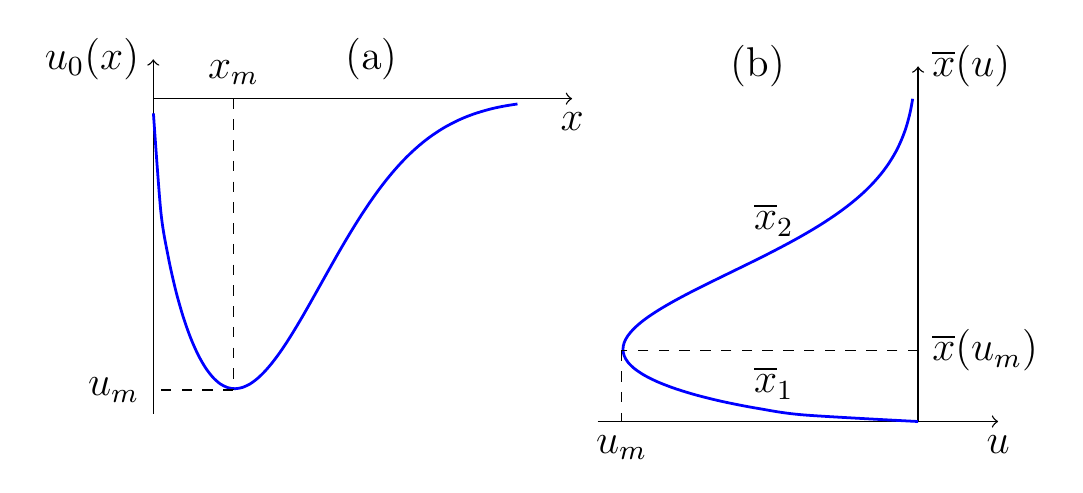}}
\caption{(a) Initial profile of a negative localized pulse. (b) Inverse function is represented by
two branches $F_1(u)$ and $F_2(u)$.
}
\label{fig7}
\end{figure}

Generalization of this calculation on localized pulses (see Fig.~\ref{fig7})
is straightforward and we present here
the final results only for $t>t_m=t(u_m)$:
\begin{equation}\label{eq66}
\begin{split}
  t(u)=&\frac1{4(-u)^{3/2}}\int_0^{u_m}\sqrt{-u}\,{\ox'_1(u)}\,du\\
  &+\frac1{4(-u)^{3/2}}\int_{u_m}^u\sqrt{-u}\,{\ox'_2(u)}\,du\\
  x_R(u)=&-\frac3{2(-u)^{1/2}}\int_0^{u_m}\sqrt{-u}\,{\ox'_1(u)}\,du\\
  &-\frac3{2(-u)^{1/2}}\int_{u_m}^u\sqrt{-u}\,{\ox'_2(u)}\,du+\ox_2(u).
  \end{split}
\end{equation}
For asymptotically large time, when $|u|\ll|u_m|$ at the soliton edge, we get
\begin{equation}\label{eq68}
  x_R\approx-\frac3{2^{1/3}}A^{2/3}t^{1/3},
\end{equation}
where
\begin{equation}\label{eq69}
  A=\int_0^{\infty}\sqrt{-u_0(x)}\,dx.
\end{equation}

The number of waves conservation law (\ref{eq1}) can be also reduced to the ordinary
differential equation (\ref{eq37}) and it should be solved with the boundary condition
that $k=0$ at the location of the soliton edge. However, this solution can be found up
to the moment $t_m$ only when the soliton edge reaches the minimum $u_m$ of the smooth
solution. At this moment the solution is given by the formula
\begin{equation}\label{eq70}
  k=2\sqrt{u-u_m}
\end{equation}
which gives the spectrum of wave numbers generated at the small-amplitude edge. The
maximal wave number corresponds to $u=0$ and is given by $k_m=2\sqrt{-u_m}$.
Consequently, the small-amplitude edge propagates at asymptotically large time
with the group velocity
\begin{equation}\label{eq71}
  \frac{dx_L}{dt}=\left.\frac{d\om}{dk}\right|_{k=k_m}=12u_m.
\end{equation}
This estimate is confirmed by numerical simulations.

In the above calculations we did not use the fact of complete integrability
of the KdV equation, hence our method can be easily applied to other equations
with known harmonic and soliton dispersion laws $\om(k)$ and $\tom(\tk)$.

\subsubsection{Generalized KdV equation: Positive pulse}

To illustrate the application of our approach to DSWs whose evolution obeys non-integrable
wave equations, we shall turn to the generalized KdV equation
\begin{equation}\label{eq72}
  u_t+V(u)u_x+u_{xxx}=0,
\end{equation}
where we suppose that the function $V(u)$, $V(0)=0$, increases monotonously with growth
of $u$ which excludes complications arising in the case of non-genuinely nonlinear
situations (see, e.g., \cite{kamch-12}). Other conditions of existence of periodic solutions
of this equations and applicability of the Whitham theory of modulations are indicated in
Ref.~\cite{el-05}. The solution of the dispersionless Hopf equation
\begin{equation}\label{eq73}
  u_t+V(u)u_x=0
\end{equation}
is given in implicit form by the formula
\begin{equation}\label{eq74}
  x-V(u)t=\ox(u),
\end{equation}
where $\ox(u)$ is the function inverse to the initial distribution $u=u_0(x)$.
[Its non-monotonous version is shown in Fig.~\ref{fig6}(b).]
We assume that the solution breaks at the moment $t=0$ which
imposes the condition $F(u)/V(u)\to0$ for $|u|\to0$.
The smooth solution matches the small-amplitude edge
at $x=x_L(t)$ of the DSW and our task now is to find this function $x_L(t)$.
Proceeding in the same way as in the KdV equation case, we use the number
of waves conservation law
\begin{equation}\label{eq75}
  \frac{\prt k}{\prt t}+\frac{\prt \om}{\prt x}=0,\qquad
  \om(\overline{u},k)=V(\overline{u})k-k^3
\end{equation}
with $k=k(\ou)$ to reduce it to the differential equation
\begin{equation}\label{eq76}
  3k\frac{dk}{d\ou}=\frac{dV(\ou)}{d\ou}
\end{equation}
and, extrapolating it on the whole DSW, we solve it with the boundary condition
$k(0)=0$ to obtain the wave number $k$ at $x=x_L$,
\begin{equation}\label{eq77}
  k=\sqrt{2V(u)/3}.
\end{equation}
Consequently, the velocity of this edge as a function of the value of $u$ at $x=x_L$ is given by
\begin{equation}\label{eq78}
  v_-=\frac{d\om}{dk}=-V(u),
\end{equation}
and it corresponds to the characteristic velocity of the limiting Whitham equation
\begin{equation}\label{eq79}
  \frac{\prt x_L}{\prt u}+V(u)\frac{\prt t}{\prt u}=0.
\end{equation}
The compatibility condition of this equation with Eq.~(\ref{eq74}) leads to the
differential equation
\begin{equation}\label{eq80}
  2V(u)\frac{dt}{du}+\frac{dV}{du}\,t=-\frac{d\ox}{du}
\end{equation}
whose solution with the initial condition $t(0)=0$ reads
\begin{equation}\label{eq81}
  t(u)=-\frac1{2\sqrt{V(u)}}\int_0^u\frac{\ox'(u)}{\sqrt{V(u)}}\,du
\end{equation}
and, consequently,
\begin{equation}\label{eq72}
  x(u)=-\frac{\sqrt{V(u)}}2\int_0^u\frac{\ox'(u)}{\sqrt{V(u)}}\,du+\ox(u).
\end{equation}
Generalization to localized initial pulses can be done in close analogy with
the above considered KdV equation case and it yields the formulas
\begin{equation}\label{eq73}
\begin{split}
  t(u)=&-\frac1{2\sqrt{V(u)}}\int_0^{u_m}\frac{\ox'_1(u)}{\sqrt{V(u)}}\,du\\
  &-\frac1{2\sqrt{V(u)}}\int_{u_m}^u\frac{\ox'_2(u)}{\sqrt{V(u)}}\,du\\
  x_L(u)=&-\frac{\sqrt{V(u)}}2\int_0^{u_m}\frac{\ox'_1(u)}{\sqrt{V(u)}}\,du\\
  &-\frac{\sqrt{V(u)}}2\int_{u_m}^u\frac{\ox'_2(u)}{\sqrt{V(u)}}\,du+\ox_2(u)
  \end{split}
\end{equation}
with obvious notation (see Fig.~\ref{fig6}). These formulas define the function $x_L(t)$
in parametric form.

Again at the asymptotically large time this pulse evolves into the sequence of solitons
whose number can be calculated by the method of Refs.~\cite{egs-08,egkkk-07}. Velocity
of the leading soliton can be found with the use of Eq.~(\ref{eq4}) in vicinity of the
moment when the small-amplitude edge reaches the point with $u=u_m$. This equation with
$\tk=\tk(\ou)$ reduces to
\begin{equation}\label{eq73b}
  3\tk\frac{d\tk}{d\ou}=-\frac{dV}{d\ou}
\end{equation}
and it should be solved with the boundary condition $\tk=0$ for $u=u_m$ which gives
\begin{equation}\label{eq73c}
  \tk=\sqrt{2[V(u_m)-V(u)]/3}.
\end{equation}
Since $V(u)$ is a monotonously increasing with $u$ function, we get the maximal value
of $\tk$, $\tk_m=\sqrt{2V(u_m)/3}$, and, consequently, the maximal soliton's velocity,
which is equal to the asymptotic velocity of the soliton edge, is given by the formula
\begin{equation}\label{eq73c}
  \frac{dx_R}{dt}=\frac{\om(\tk_m)}{\tk_m}=\tk_m^2=\frac23V(u_m).
\end{equation}

\subsubsection{Generalized KdV equation: Negative pulse}

Now we turn to a situation shown in Fig.~\ref{fig7} with the smooth solution of the
dispersionless equation given by
\begin{equation}\label{eq85}
  x-V(u)t=\ox(u).
\end{equation}
At first we consider a monotonous initial pulse.
At the soliton edge the equation
\begin{equation}\label{eq86}
  \frac{\prt \tk}{\prt t}+\frac{\prt \tom(\tk)}{\prt x}=0,\quad \tom(\ou,\tk)=V(\ou)\tk+\tk^3
\end{equation}
can be reduced to the equation
\begin{equation}\label{eq87}
  3\tk\frac{d\tk}{d\ou}=-\frac{dV}{d\ou}
\end{equation}
whose extrapolation to the whole DSW followed by solving it with the initial condition
$\tk(0)=0$ gives
\begin{equation}\label{eq88}
  \tk=\sqrt{-2V(u)/3}.
\end{equation}
Hence the soliton edge propagates with velocity
\begin{equation}\label{eq89}
  v_+=\frac{\tom}{\tk}=\frac13V(u)
\end{equation}
which coincides with the characteristic velocity of the limiting Whitham equation
\begin{equation}\label{eq90}
  \frac{\prt x_R}{\prt u}-\frac13V(u)\frac{\prt t}{\prt u}=0.
\end{equation}
Its compatibility condition with Eq.~(\ref{eq85}) leads to the differential equation
\begin{equation}\label{eq91}
  \frac23V(u)\frac{dt}{du}+\frac{dV}{du}=-\frac{d\ox}{du}
\end{equation}
whose solution with the initial condition $t(0)=0$ yields
\begin{equation}\label{eq92}
\begin{split}
  t(u)=&\frac3{2(-V(u))^{3/2}}\int_0^u {\sqrt{-V(u)}}\,{\ox'(u)}\,du\\
   x_R(u)=&-\frac3{2\sqrt{-V(u)}}\int_0^u {\sqrt{-V(u)}}\,{\ox'(u)}\,du+\ox(u).
  \end{split}
\end{equation}
For a localized pulse we obtain the formulas
\begin{equation}\label{eq94}
\begin{split}
  t(u)=&\frac3{2(-V(u))^{3/2}}\int_0^{u_m} {\sqrt{-V(u)}}\,{\ox'_1(u)}\,du\\
  &+\frac3{2(-V(u))^{3/2}}\int_{u_m}^u{\sqrt{-V(u)}}\,{\ox'_1(u)}\,du\\
  x_R(u)=&-\frac3{2\sqrt{-V(u)}}\int_0^{u_m}{\sqrt{-V(u)}}\,{\ox'_1(u)}\,du\\
  &-\frac3{2\sqrt{-V(u)}}\int_{u_m}^u{\sqrt{-V(u)}}\,{\ox'_1(u)}\,du+\ox_2(u)
  \end{split}
\end{equation}
which define in a parametric form the law of motion of the soliton edge.
For asymptotically large $t$ we get
\begin{equation}\label{eq97}
  x_R\approx-\left(\frac{3A}2\right)^{2/3}t^{1/3},\qquad
  A=\int_0^{\infty}\sqrt{-V(u_0(x))}\,dx.
\end{equation}

The spectrum of wave numbers generated at the small-amplitude edge can be
found by solving the appropriate reduction of Eq.~(\ref{eq1}) which yields
\begin{equation}\label{eq95}
  k=\sqrt{2[V(u)-V(u_m)]/3}
\end{equation}
and the small-amplitude edge propagates with the maximal group velocity
\begin{equation}\label{eq96}
  \frac{dx_L}{dt}=\left.\frac{d\om}{dk}\right|_{k=k_m}=2V(u_m),
\end{equation}
where $k_m=\sqrt{-2V(u_m)/3}$.

The examples considered here demonstrate clearly enough how to apply the
method to non-integrable wave equations in the case of unidirectional propagation.

\subsection{Bidirectional propagation}\label{sec3b}

We consider here typical situations when the wave is described by the two variables,
say, by the density $\rho$ and the flow velocity $u$, and the main supposition is
that in the dispersionless limit the equations can be transformed to the Riemann
diagonal form
\begin{equation}\label{eq98}
\begin{split}
  & \frac{\prt r_+}{\prt t}+v_+(r_+,r_-)\frac{\prt r_+}{\prt x}=0,\\
  & \frac{\prt r_-}{\prt t}+v_-(r_+,r_-)\frac{\prt r_-}{\prt x}=0
  \end{split}
\end{equation}
for the Riemann invariants $r_{\pm}$. We assume also that at the wave breaking moment
one of the Riemann invariants can be considered as constant. In many typical situations,
when the initial pulse splits into two well separated pulses corresponding to
different characteristic velocities $v_{\pm}$, this assumption is fulfilled by
virtue of the wave dynamics and we can consider wave breaking of simple waves only.
Besides that, we confine ourselves to situations when this simple wave propagates
into uniform quiescent medium so that the arising DSW belongs in integrable cases
to the class of quasi-simple waves of Ref.~\cite{gkm-89}. To compare the results
obtained by our method with known formulas derived with the use of the inverse
scattering transform method, we shall consider first the integrable NLS equation
(\ref{eq19}), but we are going to treat it without use of its complete integrability.

\subsubsection{NLS equation: Positive profile of $r_+$-invariant}

We shall start with the situation shown symbolically in Fig.~\ref{fig5}(a,b)
with the breaking of the invariant
$r_+=u/2+\sqrt{\rho}$, so that
\begin{equation}\label{eq99}
  r_-=\frac{u}2-\sqrt{\rho}=-\sqrt{\rho_0},
\end{equation}
where $\rho_0$ is the density of the quiescent medium into which the pulse propagates.
Since $\rho$ and $u$ are related by Eq.~(\ref{eq99}), it is convenient to consider
them as functions of some other variable and, as we shall see, it is convenient for
further generalizations to choose the local sound velocity, equal
in our present case to $c=\sqrt{\rho}$, as such a variable. Then $u=2(c-c_0)$, $c_0=\sqrt{\rho_0}$,
$r_+=2c-c_0$, so that the solution of dispersionless equations can be written as
\begin{equation}\label{eq100}
  x-(3c-2c_0)t=\ox(c-c_0),
\end{equation}
where $\ox(c-c_0)$ is the function inverse to the initial distribution of the local
sound velocity which we shall write in the form $c(x)=c_0+\tilde{c}_0(x)$, where
$\tilde{c}$ denotes a deviation from the background sound velocity $c_0$.
Thus, we consider the wave breaking in terms of the local sound velocity.

At the boundary with the smooth solution we have now the soliton edge $x_L$ of the
DSW and the wave breaking diagrams in Fig.~\ref{fig5}(a,b) correspond, according to our
conjecture, to the applicability conditions of Eq.~(\ref{eq4}),
\begin{equation}\label{eq101}
  \frac{\prt \tk}{\prt t}+\frac{\prt \tom(\tk)}{\prt x}=0,\quad
  \tom=\overline{u}\tk+\tk\sqrt{\overline{\rho}-\tk^2/4},
\end{equation}
where $\overline{\rho},\overline{u}$ are the density and the flow velocity of the
background along which solitons propagate. In vicinity of the soliton edge we can
consider $\tk$ and $\overline{\rho}$ as functions of $c=\sqrt{\overline{\rho}}$
and introduce a new variable
\begin{equation}\label{eq102}
  \tal(c)=\sqrt{1-\frac{\tk^2(c)}{4c^2}},
\end{equation}
so that
\begin{equation}\label{eq103}
\begin{split}
  \tk(c)&=2c\sqrt{1-\tal^2},\\
  \tom(c)&=2c\sqrt{1-\tal^2}[2(c-c_0)+c\tal].
  \end{split}
\end{equation}
With the use of the equation
\begin{equation}\label{eq104}
  \frac{\prt c}{\prt t}+(u+c)\frac{\prt c}{\prt x}=0,
\end{equation}
corresponding to one of the limiting characteristic velocities of the Whitham system
and equivalent to Eq.~(\ref{eq98}) for $r_+$, we reduce Eq.~(\ref{eq101}) to the
ordinary differential equation (see the Appendix for more details)
\begin{equation}\label{eq105}
  \frac{d\tal}{dc}=-\frac{1+\tal}c.
\end{equation}
Following El \cite{el-05}, we solve it with the
boundary condition
\begin{equation}\label{eq106}
  \tal(c_0)=1
\end{equation}
which in our `extrapolation' interpretation means that at the small-amplitude edge
the inverse width $\tk$ of solitons
vanishes together with their amplitude [see Eq.~(\ref{eq27})]. As a result, we get
\begin{equation}\label{eq107}
  \tal(c)=\frac{2c_0}c-1
\end{equation}
and
\begin{equation}\label{eq108}
  V_s=\tom/\tk=c.
\end{equation}
It corresponds to another limiting characteristic velocity of the Whitham system
and the corresponding Whitham equation can be written in the form
\begin{equation}\label{eq109}
  \frac{\prt x_L}{\prt c}-c\frac{\prt t}{\prt c}=0.
\end{equation}
The compatibility condition of this equation with Eq.~(\ref{eq100}) gives with account
of Eq.~(\ref{eq104})
\begin{equation}\label{eq110}
  2z\frac{dt}{dz}+3t=-\frac{d\ox}{dz},\quad z=c-c_0.
\end{equation}
Its solution with the initial condition $t(0)=0$ reads
\begin{equation}\label{eq111}
  t(z)=-\frac1{2z^{3/2}}\int_0^z\sqrt{z}\ox'(z)dz
\end{equation}
and, consequently,
\begin{equation}\label{eq112}
  x_L(z)=(3z+c_0)t(z)+\ox(z).
\end{equation}
For example, for the initial profile $\ox(z)=-z^n$ we get
\begin{equation}\label{eq112b}
  x_L(t)=c_0t+\frac{n-1}{2n}\left(1+\frac1{2n}\right)^{\frac1{n-1}}t^{\frac{n}{n-1}}.
\end{equation}
This formula coincides, up to the notation, with the result of Ref.~\cite{kamch-18} obtained
from the global solution of the Whitham equations with the use of complete integrability
of the NLS equation. Up to the notation, the formulas (\ref{eq111}) and (\ref{eq112})
reproduce the law of motion of the soliton edge derived in Ref.~\cite{ekkag-09} in a
different physical context of the flow of Bose-Einstein condensate past a wing.
All that confirms the validity of our approach.

Generalization of these formulas to the case of localized pulses is straightforward,
\begin{equation}\label{eq111b}
  \begin{split}
   & t(z)=-\frac1{2z^{3/2}}\left\{\int_0^{z_m}\sqrt{z}\ox'_1(z)dz-
   \int_{z_m}^z\sqrt{z}\ox'_2(z)dz\right\},\\
  & x_L(z)=(3z+c_0)t(z)+\ox_2(z).
  \end{split}
\end{equation}
For asymptotically large time we find
\begin{equation}\label{eq113}
\begin{split}
  & x_L=c_0t+\left(\frac{3A}2\right)^{2/3}t^{1/3},\\
  & A=\int_{-\infty}^0\sqrt{\widetilde{c}_0(x)}\,dx,
  \end{split}
\end{equation}
where $c=c_0+\widetilde{c}_0(x)$ is the initial distribution of the local sound velocity.

To find the velocity of the small-amplitude edge, we have to solve the number
of waves conservation law equation
\begin{equation}\label{eq115}
  \frac{\prt k}{\prt t}+\frac{\prt\om}{\prt x}=0,\quad
  \om=uk+k\sqrt{c^2+\frac{k^2}4}.
\end{equation}
Under the same assumptions as above, it can be reduced to the equation (see Appendix)
\begin{equation}\label{eq116}
  \frac{d\al}{dc}=-\frac{1+\al}c
\end{equation}
for the variable
\begin{equation}\label{eq117}
  \al(c)=\sqrt{1+\frac{k^2(c)}{4c^2}},
\end{equation}
so that
\begin{equation}\label{eq118}
\begin{split}
  & k(c)=2c\sqrt{\al^2-1},\\
  & \om(c)=2c\sqrt{\al^2-1}[2(c-c_0)+c\al].
  \end{split}
\end{equation}
Now Eq.~(\ref{eq116}) must be solved with the boundary condition that when
the soliton edge reaches the point where $c=c_m$, we have $k=0$ and $\al(c_m)=1$.
This gives
\begin{equation}\label{eq119}
  \al(c)=\frac{2c_m}c-1,\quad c_0\leq c\leq c_m,
\end{equation}
and $k(c)=4\sqrt{c_m(c_m-c)}$. Consequently, the range of wave numbers generated at the
small-amplitude edge is given by
\begin{equation}\label{eq120}
  0\leq k\leq k_m=4\sqrt{c_m(c_m-c_0)},
\end{equation}
and it is easy to find that the maximal group velocity is equal to
\begin{equation}\label{eq121}
  \frac{dx_R}{dt}=\left.\frac{d\om}{dk}\right|_{k=k_m}=2r_m-\frac{c_0^2}{r_m}
\end{equation}
where $r_m=2c_m-c_0$ is the maximum value of $r_+$ in its initial distribution.
If $c_0\ll r_m$, then the expression for the velocity of the small-amplitude edge simplifies to
\begin{equation}\label{eq122}
  \frac{dx_R}{dt}\approx 2r_m.
\end{equation}

Thus, we have found the laws of motion of both edges of the DSW for the case of breaking
of the dispersionless invariant $r_+$. DSW generated after breaking of the wave
propagating in the opposite direction can be considered in a similar
way and the resulting formulae differ from the above ones by obvious changes of some signs.

\subsubsection{NLS equation: Negative profile of $r_+$-invariant}

Now we turn to the situation shown symbolically in Figs.~\ref{fig5}(c,d) with such
breaking of the invariant $r_+$
that the smooth solution matches the small-amplitude edge of the DSW and in the
integrable approach both Riemann invariants $r_2$ and $r_3$ change along the DSW.
At first we shall find the law
of motion of the small amplitude edge.

We represent the smooth solution of the dispersionless equation in the form
\begin{equation}\label{eq123}
  x-(3c-2c_0)t=\ox(c-c_0),
\end{equation}
where $\ox(z)>0$ for $z<0$. Equation (\ref{eq1}) can be reduced, as in the preceding section,
to Eq.~(\ref{eq115}) whose solution is looked for with the boundary condition $\al(c_0)=1$
at the soliton edge which gives
\begin{equation}\label{eq124}
  \al(c)=\frac{2c_0}c-1,\quad 0<c<c_0,
\end{equation}
so that $k=4\sqrt{c_0(c_0-c)}$. Then the group velocity is equal to
\begin{equation}\label{eq125}
  v_-=\frac{d\om}{dk}=2c_0-\frac{c^2}{2c_0-c},
\end{equation}
and it corresponds to the characteristic velocity of the limiting Whitham equation
\begin{equation}\label{eq126}
  \frac{\prt x}{\prt c}-\left(2c_0-\frac{c^2}{2c_0-c}\right)\frac{\prt t}{\prt c}=0.
\end{equation}
Its compatibility condition with Eq.~(\ref{eq123}) leads to the differential equation
\begin{equation}\label{eq127}
  \frac{(4c_0-c)(c_0-c)}{2c_0-c}\frac{dt}{dc}-\frac32t=\frac12\ox'(c-c_0)
\end{equation}
whose solution reads
\begin{equation}\label{eq128}
\begin{split}
  &t(c)=\frac1{2(4c_0-c)\sqrt{c_0-c}}\int_{c_0}^c\frac{(2c_0-c)\ox'(c-c_0)}{\sqrt{c_0-c}}dc,\\
  &x_R(c)=(3c-2c_0)t(c)+\ox(c).
  \end{split}
\end{equation}
In the case of a localized pulse we obtain
\begin{equation}\label{eq130}
\begin{split}
  t(c)=&\frac1{2(4c_0-c)\sqrt{c_0-c}}\Big\{\int_{c_0}^c\frac{(2c_0-c)\ox'_1(c-c_0)}{\sqrt{c_0-c}}dc\\
  &+\int_c^{c_0}\frac{(2c_0-c)\ox'_2(c-c_0)}{\sqrt{c_0-c}}dc\Big\},\\
  x_R(c)&=(3c-2c_0)t(c)+\ox_2(c).
  \end{split}
\end{equation}
Such a pulse evolves eventually to a sequence of dark solitons and their number can be found with the
use of the Bohr-Sommerfeld quantization rule.

For finding the trailing soliton velocity at asymptotically large time we solve Eq.~(\ref{eq105})
with the boundary condition $\tk=0$ at $c=c_m$ or $\tal(c_m)=1$. This gives
\begin{equation}\label{eq130b}
  \tal(c_0)=\frac{2c_m}{c_0}-1,
\end{equation}
and consequently the trailing soliton velocity is equal to
\begin{equation}\label{eq130c}
  \frac{dx_L}{dt}={\tom}/{\tk}=c_0\tal(c_0)=2c_m-c_0=r_m.
\end{equation}
We can compare this with the result of the Bohr-Sommerfeld quantization rule (see, e.g., \cite{kku-02})
which gives the expression of the soliton velocity in terms of the Riemann invariants $r_i,$
$i=1,2,3,4,$ of the Whitham equations, $V_s=\tfrac12\sum_ir_i$. In the case of the present
initial conditions they are equal to $r_1=-\sqrt{\rho_0}$, $r_2=r_3=r_m$, $r_4=\sqrt{\rho_0}$,
and we obtain $V_s$ coinciding with (\ref{eq130c}).

\subsubsection{Generalized NLS equation: Positive profile of $r_+$-invariant}

To illustrate application of the method to non-integrable equations, we shall consider
the generalized NLS equation
\begin{equation}\label{eq131}
    i\psi_t+\frac12\psi_{xx}-f(|\psi|^2)\psi=0
\end{equation}
where the nonlinearity function $f(\rho)$, $f(0)=0$, is supposed to be increasing positive function of
the density $\rho=|\psi|^2$. Linear harmonic waves propagating along background with density $\rho$
satisfy the dispersion law (\ref{eq115}) where
\begin{equation}\label{eq133}
  c=\sqrt{\rho f'(\rho)}
\end{equation}
is the sound velocity (see, e.g., \cite{ks-09}). In the dispersionless limit the wave dynamics equations
can be written in Riemann form (\ref{eq98}) with the Riemann invariants
\begin{equation}\label{eq134}
  r_{\pm}=\frac{u}2\pm\frac12\int_0^{\rho}\frac{cd\rho}{\rho}
\end{equation}
and with the characteristic velocities
\begin{equation}\label{eq134b}
  v_{\pm}=u\pm c.
\end{equation}
It is convenient to replace the density $\rho$ as a wave variable by the sound velocity $c$, so that
$\rho=\rho(c)$ is the function inverse to $c=c(\rho)$ defined in Eq.~(\ref{eq133}) and the Riemann
invariants take the form
\begin{equation}\label{eq135}
  r_{\pm}=\frac{u}2\pm\sigma(c),\quad \sigma(c)=\frac12\int_0^c\frac{c\rho'(c)}{\rho(c)}dc.
\end{equation}
Evolution of initial step-like distributions was studied in much detail in Ref.~\cite{hoefer-14}
and we turn here to the problem of evolution of non-uniform initial distributions of
simple wave type.

In this section we consider wave breaking of a simple wave with constant Riemann invariant
\begin{equation}\label{eq136}
\begin{split}
  &r_-=\frac{u}2-\sigma(c)=-\sigma(c_0),\\
  &\text{i.e.}\quad u=2[\sigma(c)-\sigma(c_0)].
  \end{split}
\end{equation}
This simple wave propagates into a quiescent medium where the constant sound velocity
equals $c_0$.
Its evolution equation can be written in the form obtained from the first equation
(\ref{eq98}) with $r_+=2\sigma(c)-\sigma(c_0)$,
\begin{equation}\label{eq137}
  \frac{\prt c}{\prt t}+(c+u)\frac{\prt c}{\prt x}=0,
\end{equation}
whose solution reads
\begin{equation}\label{eq138}
  x-\left\{2[\sigma(c)-\sigma(c_0)]+c\right\}t=\ox(c-c_0),
\end{equation}
where we suppose that the initial distribution of $c$ is positive, i.e., $\ox(c-c_0)<0$ for $c>c_0$
and the inverse function $\widetilde{c}_0(x)=\ox^{-1}(x)$ defines the initial distribution of
the local sound velocity $c_0+\widetilde{c}_0(x)$ such that the wave breaks at the moment $t=0$.
The corresponding breaking pattern shown symbolically in Figs.~\ref{fig5}(a,b) leads to the
formation of DSW with the soliton edge at $x_L$. Since in this case Eq.~(\ref{eq4}) is applicable,
we can find the law of motion of this edge by our method.

We write the `soliton dispersion law' in the form (see Eqs.~(\ref{eq101}) and (\ref{eq102}))
\begin{equation}\label{eq139}
  \tom=\tk(u+c\tal(c))
\end{equation}
and its substitution together with $\tk(c)=2c\sqrt{1-\tal^2}$ gives with account of
Eqs.~(\ref{eq136}) and (\ref{eq137}) the differential equation (see the Appendix)
\begin{equation}\label{eq140}
  \frac{d\tal}{dc}=-\frac{(1+\tal)(2\sigma'(c)+2\tal-1)}{c(1+2\tal)},
\end{equation}
which, according to the extrapolation rule, should be solved with the boundary condition
$\tal(c_0)=1$, that is the soliton inverse width vanishes together with its amplitude
at the small-amplitude edge.

Generally speaking, Eq.~(\ref{eq140}) can be solved only numerically, but if
$\sigma'(c)=1/p=\mathrm{const}$, that is
\begin{equation}\label{eq141}
  f(\rho)=\frac1p\rho^p\quad\text{and}\quad c=\rho^{p/2},
\end{equation}
then an easy calculation gives
\begin{equation}\label{eq142}
  c(\tal)=c_0\left(\frac2{1+\tal}\right)^{\frac{p}{3p-1}}
  \left(\frac{2+p}{2-p+2p\tal(c)}\right)^{\frac{2(p-1)}{3p-2}}.
\end{equation}
Naturally, for $p=1$ this formula reproduces the result (\ref{eq107}) of the
NLS equation theory. The function $c(\tal)$ can be also inverted for $p=2$,
\begin{equation}\label{eq143}
  \tal(c)=\frac12\left(\sqrt{1+8\left(\frac{c_0}c\right)^2}-1\right),
\end{equation}
and for $p=1/2$,
\begin{equation}\label{eq144}
  \tal(c)=\frac1{16}\left\{\sqrt{\frac{c_0}{c}\left(25\frac{c_0}{c}-16\right)}
  +25\frac{c_0}{c}-24\right\}.
\end{equation}
Up to the notation, these formulas coincide with those found in Ref.~\cite{hoefer-14}.

When the function $\tal(c)$ is known, we can find the law of motion of the soliton edge.
Indeed, the soliton velocity
\begin{equation}\label{eq145}
  V_s=\tom/\tk=2[\sigma(c)-\sigma(c_0)]+c\tal(c)
\end{equation}
corresponds to the characteristic velocity of the limiting Whitham equation
\begin{equation}\label{eq146}
  \frac{\prt x}{\prt c}-\frac{\tom}{\tk}\frac{\prt t}{\prt c}=0,
\end{equation}
which must be compatible with Eq.~(\ref{eq138}) and this condition yields the equation
\begin{equation}\label{eq147}
  c(1-\tal(c))\frac{dt}{dc}+(1+2\sigma'(c))t=-\ox'(c-c_0).
\end{equation}
Its solution with the initial condition $t(c_0)=0$ reads
\begin{equation}\label{eq148}
\begin{split}
  t(c)=-G(c)
  \int_{c_0}^c \frac{\ox'(c-c_0)dc}{c(1-\tal(c))G(c)}
  \end{split}
\end{equation}
and, consequently,
\begin{equation}\label{eq149}
  x_L(c)=\left\{2[\sigma(c)-\sigma(c_0)]+c\right\}t(c)+\ox(c-c_0),
\end{equation}
where
\begin{equation}\label{eq149b}
  G(c)=\exp\left\{-\int_{a}^c\frac{(1+2\sigma'(c))dc}{c(1-\tal(c))}\right\},
\end{equation}
the integration limit $a$ is chosen so that the integral is convergent. In fact,
the functions $t(c),\,x_L(c)$ do not depend on $a$. In particular, for $f(\rho)$
given by Eq.~(\ref{eq141}) we obtain up to inessential constant factor
\begin{equation}\label{eq149c}
  G(c)=\frac{[1+\tal(c)]^{\frac{p+2}{2(3p-2)}}[2p\tal(c)+2-p]^{\frac{4(p-1)}{3p-2}}}
  {[1-\tal(c)]^{3/2}}.
\end{equation}
Generalization on localized pulses is straightforward and we shall not write down
quite lengthy formulas.

As in the case of the standard NLS equation, we can find velocity of the small-amplitude
edge at asymptotically large time for a localized initial pulse by solving
Eq.~(\ref{eq115}) provided that the flow velocity
$u$ is given by Eq.~(\ref{eq136}). Then this equation reduces again to Eq.~(\ref{eq147}),
but now it should be solved with the boundary condition
\begin{equation}\label{eq150}
  \al(c_m)=1,
\end{equation}
where $c_m$ is the maximal value of the local velocity in the initial distribution
$c(x)=c_0+\widetilde{c}_0(x)$. If $f(\rho)=\rho^p/p$, then the solution is given by
Eq.~(\ref{eq142}) or by its particular cases for $p=1,2,1/2$ with $c_0$ replaced by $c_m$.
The spectrum of wave numbers $k=2c\sqrt{\al^2(c)-1}$ with $c$ in the range
$c_0\leq c\leq c_m$ has the maximal value $k_m=2c_0\sqrt{\al^2(c_0)-1}$ and the
corresponding group velocity
\begin{equation}\label{eq151}
  \frac{dx_R}{dt}=\left.\frac{d\om}{dt}\right|_{k=k_m}=2c_0\al(c_0)-\frac{c_0}{\al(c_0)}
\end{equation}
equals the asymptotic velocity of the small-amplitude edge. If $\al(c_0)=2c_m/c_0-1$,
then this formula reproduces the known result (\ref{eq121}).

\subsubsection{Generalized NLS equation: Negative profile of $r_+$-invariant}

If the initial profile of local sound velocity $c(x)=c_0+\widetilde{c}_0(x)$ has the
form of a ``hole'' $\widetilde{c}_0(x)<0$, then we represent the smooth dispersionless
solution by the formula
\begin{equation}\label{eq152}
  x-\left\{2[\sigma(c)-\sigma(c_0)]+c\right\}t=\ox(c-c_0),
\end{equation}
where $\ox(c-c_0)>0$ for $c<c_0$, that is the initial distribution of $c$ differs from $c_0$
at $x>0$ only. Equation (\ref{eq1}) reduces again to Eq.~(\ref{eq140}) and its solution with
the boundary condition $\al(c_0)=1$ coincides with Eq.~(\ref{eq142}) where $\tal$ is replaced
by $\al$ and now we have $\al>1$. Correspondingly, in formulas for particular cases $p=1,2,1/2$ (see
Eqs.~(\ref{eq107}), (\ref{eq143}), (\ref{eq144})) we should assume $c<c_0$.

The small-amplitude edge propagates with the group-velocity
\begin{equation}\label{eq153}
  \frac{dx_L}{dt}=\frac{d\om}{dk}=2[\sigma(c)-\sigma(c_0)+c\al(c)]-\frac{c}{\al(c)},
\end{equation}
which can be considered as a characteristic velocity of the limiting Whitham equation
\begin{equation}\label{eq154}
  \frac{\prt x}{\prt c}-\left\{2[\sigma(c)-\sigma(c_0)+c\al(c)]-\frac{c}{\al(c)}\right\}
  \frac{\prt t}{\prt c}=0.
\end{equation}
Its compatibility condition with Eq.~(\ref{eq149}) leads to the differential equation
\begin{equation}\label{eq155}
  \frac{c(\al(c)-1)(2\al(c)+1)}{\al(c)}\frac{dt}{dc}-(2\sigma'(c)+1)t=\ox'(c-c_0),
\end{equation}
whose solution with the initial condition $t(c_0)=0$ reads
\begin{equation}\label{eq156}
  t(c)=G(c)\int_{c_0}^c\frac{\al(c)\ox'(c-c_0)dc}{c(\al(c)-1)(2\al(c)+1)G(c)},
\end{equation}
where
\begin{equation}\label{eq157}
  G(c)=\exp\left\{\int_{a}^c
  \frac{(2\sigma'(c)+1)\al(c)dc}{c(\al(c)-1)(2\al(c)+1)}\right\},
\end{equation}
which for $f(\rho)$ given by Eq.~(\ref{eq141}) reduces up to an inessential
constant factor to
\begin{equation}\label{eq158}
  G(c)=\frac{[\al(c)+1]^{\frac{p+2}{2(3p-2)}}[2p\al(c)+2-p]^{\frac{p-2}{3p-2}}}
  {\sqrt{\al(c)-1}}.
\end{equation}
Then we get from Eq.~(\ref{eq152})
\begin{equation}\label{eq159}
  x_R(c)=\left\{2[\sigma(c)-\sigma(c_0)]+c\right\}t(c)+\ox(c-c_0).
\end{equation}
These formulas determine in a parametric form the law of motion of the
small-amplitude edge. Their generalization to localized pulses is
straightforward.

At asymptotically large time the velocity of the trailing soliton generated from
a localized pulse is determined by the value $\tal(c_0)$, where $\tal(c)$ is
the solution of Eq.~(\ref{eq140}) with the boundary condition $\tal(c_m)=1$.
Then for the velocity of the soliton edge we obtain
\begin{equation}\label{eq159b}
  \frac{dx_L}{dt} =c_0\tal(c_0)
\end{equation}
which generalizes Eq.~(\ref{eq130c}).

\section{Comparison with numerical solution}\label{sec4}

\begin{figure}[t]
\centerline{\includegraphics[width=8cm]{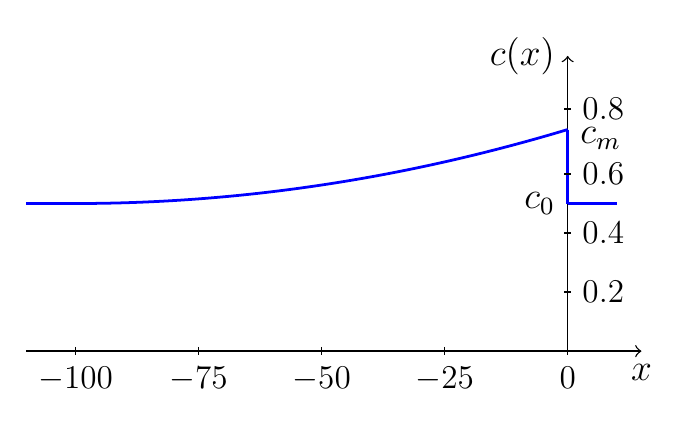}}
\caption{Initial profile of the local sound velocity used in the numerical solution
of the generalized NLS equation.
}
\label{fig8}
\end{figure}

\begin{figure}[t]
\centerline{\includegraphics[width=8cm]{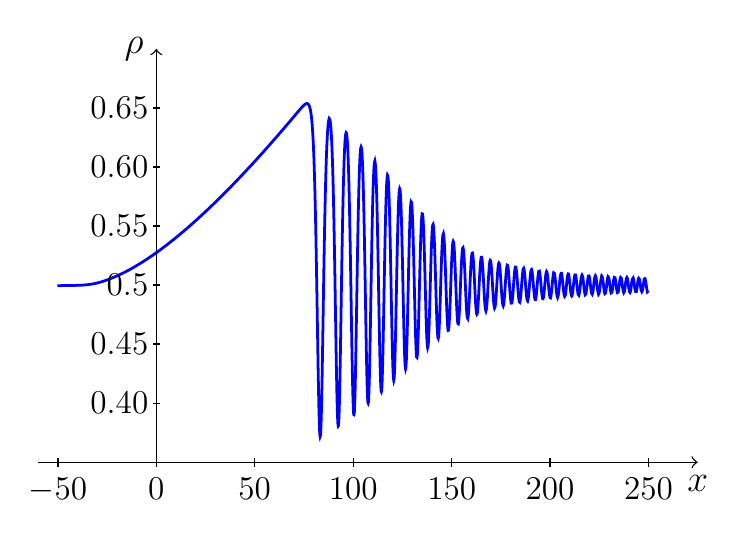}}
\caption{Plot of the density in the DSW evolved from the initial distribution of
the local sound velocity $\rho=c(x)$ given by
Eq.~(\ref{s4-1}) (for $p=2$ the density is equal to $\rho=c$ and the flow velocity
to $u(x)=c(x)-c_0$). Evolution time is $t=120$.
}
\label{fig9}
\end{figure}

\begin{figure}[t]
\centerline{\includegraphics[width=8cm]{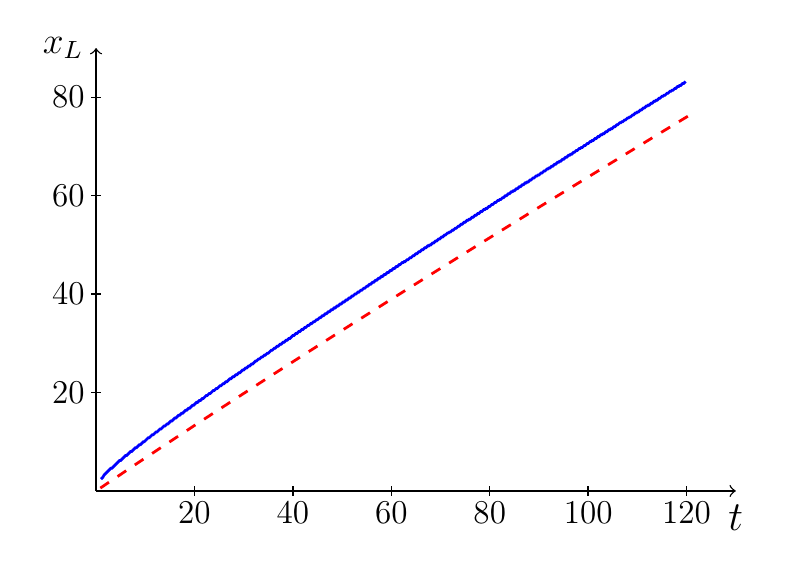}}
\caption{Law of motion of the soliton edge. Solid (blue) line shows the result of
numerical solution and dashed (red) line corresponds to the analytical formulas
(\ref{s4-4}).
}
\label{fig10}
\end{figure}

To illustrate accuracy of the method, we compare here the results of exact numerical
solution of the generalized NLS equation (\ref{eq131}) with $f(\rho)$ given by
Eq.~(\ref{eq141}) with $p=2$ for the initial local sound velocity distribution
\begin{equation}\label{s4-1}
  c(x)=
   0.5+[0.005(x+100)]^2,\quad -100\leq x\leq0,
\end{equation}
and $c(x)=0.5$ outside this interval; see Fig.~\ref{fig8}.
A typical form of the DSW generated from such a pulse is shown in Fig.~\ref{fig9}.
We have chosen the initial distribution with sharp front at $x=0$ to obtain
fast enough transition to the asymptotic regime for the small-amplitude edge of the DSW.
At the same time, large length of the pulse prevents too fast transition to the
asymptotic regime for the soliton edge and therefore the general formulas
(\ref{eq148}), (\ref{eq149}) should be used. In actual numerical solution this
sharp transition was slightly smoothed at the front edge, but we neglect here a
small contribution
of the branch $\ox_1(c-c_0)$ of the inverse function $x=\ox(c-c_0)$ and take into
account only the branch
\begin{equation}\label{s4-2}
  \ox(c-c_0)=\ox_2(c-c_0)=200\sqrt{c-c_0}-100.
\end{equation}
In our case with $p=2$ we have $\sigma(c)=c/2$, $\tal(c)$ is given by
Eq.~(\ref{eq143}), so an elementary integration in the formula (\ref{eq149b}) yields
\begin{equation}\label{s4-3}
  G(c)=\frac{c+\sqrt{c^2+8c_0^2}}{[(c+\sqrt{c^2+8c_0^2})^2-16c_0^2]^{3/2}},
\end{equation}
where we have omitted an inessential constant factor which cancels in the formulas
\begin{equation}\label{s4-4}
  \begin{split}
  & t(c)=-2G(c)\int_{c_m}^c\frac{\ox'(c-c_0)dc}{\left(3c-\sqrt{{c}^2+8c_0^2}\right)G(c)},\\
  & x_L(c)=(2c-c_0)t(c)+\ox(c-c_0).
  \end{split}
\end{equation}
Substitution of Eq.~(\ref{s4-2}) gives the parametrical dependence $x=x_L(t)$
shown in Fig.~\ref{fig10} by a dashed line. As we can see, it agrees reasonably well
without any fitting parameter with the result of numerical solution shown by a solid line.

\begin{figure}[t]
\centerline{\includegraphics[width=8cm]{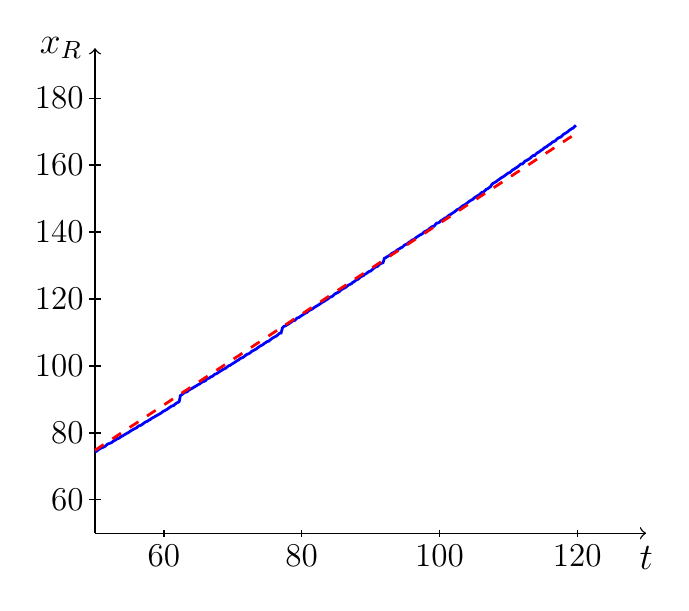}}
\caption{Linear law of motion of the small-amplitude edge with the slope calculated
according to Eq.~(\ref{eq151}) (red dashed line) and from the numerical solution
(blue solid line).
}
\label{fig11}
\end{figure}

For finding the law of motion of the small-amplitude edge, we use the formula
(\ref{eq151}) with
\begin{equation}\label{eq143b}
  \al(c_0)=\frac12\left(\sqrt{1+8\left(\frac{c_m}{c_0}\right)^2}-1\right),
\end{equation}
which gives the velocity $dx_R/dt\approx1.356$ (we take $c_m=0.742$, $c_0=0.5$ which
corresponds to the actual initial distribution). A linear dependence with this slope fits
well to the numerical solution shown in Fig.~\ref{fig11}.
This agreement should be considered as very good since the position of the
small-amplitude edge is not very certain, as is clear from Fig.~\ref{fig9},
and we determine it by means of an approximate
extrapolation of the envelopes of the wave at this edge. Some irregularities in
the numerical plot in Fig.~\ref{fig11} correspond to changes of the amplitude maxima and
minima used in such an extrapolation at different moments of time.

The above comparison of the analytical predictions with the numerical solution
demonstrates quite convincingly that the method suggested here gives accurate
enough description of pulses whose evolution obeys non-integrable equations.

\section{Conclusion}\label{sec5}

We have shown that Whitham's number of waves conservation law (\ref{eq1}) and its
soliton counterpart (\ref{eq4}) allow one to calculate the main parameters of DSWs
for a wide class of initial conditions including the pulses propagating into a medium
`at rest'. Here we formulate the main principles of this method.

(i) While Eq.~(\ref{eq1}) is universally correct in framework of the Whitham theory,
Eq.~(\ref{eq4}) has limited applicability and we have presented argumentation in
favor of its applicability to situations when the pulse under consideration propagates
into a medium at rest at some reference frame.

(ii) For a given initial distribution of the simple-wave type, the smooth solution of
the dispersionless equations can be considered as known and at its boundary with DSW
Eqs.~(\ref{eq1}) and (\ref{eq4}) reduce to ordinary differential equations which
can be extrapolated to the whole DSW, and then
their solution with known boundary condition at the opposite edge yields the wave number $k$
or the inverse half-width of solitons $\tk$ at the boundary with the smooth part of the pulse.
This procedure is equivalent to El's approach \cite{el-05}.

(iii) Consequently, the corresponding group velocity or the soliton's velocity
at the DSW edge can be expressed
in terms of the parameters of the smooth solution at its boundary with DSW.

(iv) These velocities can be treated as the characteristic velocities of the limiting
Whitham equations at this edge and this can be represented as the hodograph transformed
form of the first order partial differential equation.

(v) At last, the compatibility condition of this partial differential equation with the
smooth dispersionless solution yields the law of motion of this edge of DSW. If the soliton
solution is known, then the soliton's amplitude related with its velocity can be also found.

This scheme reproduces the known results when it is applied to the completely integrable
equations, and its applicability to non-integrable equations is confirmed by comparison
with the results of numerical simulations. Thus, the method suggested here permits one
to calculate parameters of DSWs in a number of interesting physical problems. The results
obtained here can be applied to various nonlinear optics models that reduce to different forms
of generalized NLS equation (see, e.g., \cite{kivshar-08}). Applications to shallow water
waves described by different nonlinear wave models (see, e.g. the review \cite{kdfm-18})
or to DSWs in nonlinear lattices (see, e.g., \cite{kst-97}) and in systems described by the
Gardner equation (see, e.g., \cite{sp-99}) are also of great interest.

\begin{acknowledgments}
I am grateful to S.~K.~Ivanov for help with numerical calculations. Numerous discussions of
problems of nonlinear pulses propagation with F.~Kh.~Abdullaev, G.~A.~El, M.~Isoard, S.~K.~Ivanov,
A.~I.~Maimistov, N.~Pavloff, and M.~Salerno are greatly appreciated.
The reported study was funded by RFBR according to Research Project No.~16-01-00398.

\setcounter{equation}{0}

\renewcommand{\theequation}{A.\arabic{equation}}

\section*{Appendix A. Equations for $\al(c)$ and $\tal(c)$}

Here we give for completeness some details of derivation of equations for $\al(c)$
and $\tal(c)$ directly from Eqs.~(\ref{eq1}) and (\ref{eq4}). Since in both cases the
calculations are very similar, we shall consider the equation for the small-amplitude
edge of DSW where the Whitham number of waves conservation law
\begin{equation}\label{A1}
  \frac{\prt k}{\prt t}+\frac{\prt \om}{\prt x}=0
\end{equation}
holds. We write the dispersion law in the form
\begin{equation}\label{A2}
  \om(c,k)=uk+ck\al(c,k),\quad \al(c,k)=\sqrt{1+\frac{k^2}{4c^2}},
\end{equation}
where the first term in the right-hand side represents the Doppler shift of the frequency
caused by the flow of the medium with velocity $u$ and the factor $\al(c,k)$ describes
deviation of the dispersion law from the dispersionless limit $\om=ck$, $k\to0$ $(u=0)$,
$c$ being the local sound velocity.

In dispersionless limit the generalized NLS equation
(\ref{eq131}) takes a hydrodynamic form which can be cast into equations for the Riemann
invariants (\ref{eq135}). We consider the DSW evolving from a simple wave with constant
$r_-=\mathrm{const}$ which gives according to the Gurevich-Meshcherkin conjecture \cite{gm-84}
the expression for $u$ in terms of $c$,
\begin{equation}\label{A3}
  u=2[\sigma(c)-\sigma(c_0)].
\end{equation}
Consequently,
$r_+=2\sigma(c)-\sigma(c_0)$ is a function of $c$ only and the equation for $r_+$
reduces to the equation for $c$,
\begin{equation}\label{A4}
  \frac{\prt c}{\prt t}+(u+c)\frac{\prt c}{\prt x}=0.
\end{equation}
Following El \cite{el-05}, we assume that $k$ is also a function of $c$, so substitution of
(\ref{A2}) and (\ref{A3}) into (\ref{A1}) with account of (\ref{A4}) yields after
obvious cancellations the equation
\begin{equation}\label{A5}
  \frac1k\frac{dk}{dc}\,c(1-\al)=2\sigma'+\al+c\al'.
\end{equation}
From definition of $\al$ in Eq.~(\ref{A2}) we get
\begin{equation}\label{A6}
  k(c)=2c\sqrt{\al^2(c)-1}
\end{equation}
and substitution of this expression into Eq.~(\ref{A5}) yields equation for $\al(c)$,
\begin{equation}\label{A7}
  \frac{d\al}{dc}=-\frac{(\al+1)(2\sigma'+2\al-1)}{c(2\al+1)}.
\end{equation}
The equation for $\tal(c)$ can be obtained in a similar way.

It is worth noticing that many other physical systems differ from the 
case considered here by the linear dispersion law only, that is by the concrete form of the
expression for $\al(c,k)$. If this expression is solved with respect to $k$,
then substitution of $k=k(c,\al(c))$ into Eq.~(\ref{A5}) gives the equation
for $\al(c)$ for the system under consideration.

\end{acknowledgments}

\end{document}